\documentclass[12pt]{article}
\usepackage[cp1251]{inputenc}
\usepackage[english]{babel}
\usepackage{array}
\usepackage{graphicx}
\usepackage{tabularx}
\usepackage{xltabular}
\usepackage{rotating}
\usepackage{amsmath, epsfig, amssymb}
\begin{document}


{\huge X-ray spectral features and classification of selected QSOs}

L.V. Zadorozhna$^{1}$, A.V. Tugay$^{1}$, O.I. Maluy$^{1}$ and N.G. Pulatova $^{2}$
	
	$^{1}$ Taras Shevchenko National University of Kyiv, Physical
	faculty, ave. Glushkova 2, building \textnumero 1, Kyiv, Ukraine, 03680

	$^{2}$ Main Astronomical Observatory of the National Academy of Sciences of Ukraine, Akademika Zabolotnoho str. 27, Kyiv, Ukraine, 03143

E-mail: zadorozhna\_lida@ukr.net, tugay.anatoliy@gmail.com, 

albufeira.astronomy@gmail.com, nadya@mao.kiev.ua

\begin{abstract}

We present the results of a systematic analysis of the XMM-Newton spectra of nearby optically bright QSOs. The objects have been selected from X-ray Galaxy Catalog Xgal20. It is a catalog of 1172 manually identified and classified galaxies, obtained as a cross-correlation between the 4XMM-DR9 catalog and the Hyper-Linked Extragalactic Databases and Archives (HyperLeda) with an X-ray flux greater than $F \geq 10^{-13}~\mbox{erg}/(\mbox{cm}^2\cdot\mbox{s})$. 

The goal of this work is to characterize the X-ray spectral properties of selected QSOs in the 0.1 -- 10 keV energy band. The majority of the sources (6 out of 11), are classified as radio-quiet QSOs. We studied optical spectra, hardness ratios and performed X-ray spectral fits for the 10 brighter sources. In most cases, the power law model with absorption is good enough to simulate observed continua. Although the details of the spectrum in some sources significantly complicate the model for fitting. The majority of sources have steep spectra $\Gamma > 2.1$. 
Extremely steep photon index ($\sim 2.4 - 2.5$) in our sample occurs for three radio-loud type I quasars. We detected Fe K$\alpha$ line for two radio-loud type II quasars.
We find no strong evidence for spectral hardening above 2 keV neither for quasars of type I nor for obscured type II. For each quasar its type was established both based on the features and details of observed X-ray spectrum and previous data.

\end{abstract}


\section {Introduction}

\indent \indent The most prominent extragalactic X-ray sources are the active galactic nuclei (AGN). They are the complex central structures in some galaxies, where there is an intense fall of matter onto a supermassive black hole (SMBH). The falling matter forms an accretion disk where the angular momentum of matter is converted into radiation \cite{solpiter}. Accretion process can convert about 10 percent to over 40 percent of the mass of an falling matter into energy. This central engine is surrounded by optically thick dust torus. 

According to the modern point of view the nucleus is formed together with the galaxy and grows and its current level of activity depends directly on the environment \cite{zasov}. The levels of activity of galaxies nuclei vary widely, but there are different manifestations of the same process. 
The spectrum of electromagnetic radiation of AGN is much wider than the spectrum of ordinary galaxies and can range from radio- to hard gamma- radiation. They have high luminosities: up to $10^1 \div 10^4$ times greater then normal galaxies ($L_{AGN}\sim 10^{42} - 10^{48} erg/s$). Their energy emission is mostly nonstellar, it cannot be explained as combined radiation of even trillions of stars; there should be an additional energy source to the black body radiation of galaxies \cite{peterson}. Continuum emission has almost flat spectrum from infrared to X-ray ($\nu F_{\nu}\sim  const$). Radio luminosity of AGN is 10 -- 100 times larger than for normal galaxy; X-ray emission is also $10^3 - 10^4$ times larger \cite{odo, anuradha}. 
AGN variability has characteristic time 
from 10 minutes in the X-ray range and up to 10 years in the optical and radio ranges \cite{shapovalova}, \cite{popov}. 
From spectral lines widening due to the Doppler effect one can conclude about the movement of hot gas at high speed. 
This is confirmed by visible morphological features, in particular, jets and "hot spots". AGN's radiation spectrum and its polarization signs to presence of strong magnetic fields presence and their complex structure \cite{blandford}. 

Historically, galaxies with active nuclei, where the luminosity of the nucleus does not exceed the luminosity of the host galaxy, are called Seyfert galaxies. If the luminosity of the active nucleus is much greater than the luminosity of parent galaxy, such object is called a quasar. In the optical band the conventional dividing line between Seyfert galaxies and quasars is $M_{B}=-23$ \cite{krumpe}.
Quasars (quasi-stellar radio sources) were discovered first by radio surveys. 
Their optical counterparts (QSOs, quasi-stellar objects) have distinctive emission-line spectra revealing high redshifts, strong blue continua, broad emission lines, significant X-ray flux. The terms "quasars" and "QSOs" have become interchangeable \cite{peterson}. Also the terms "radio-loud" and "radio-quiet" quasars are used to characterize their radio luminosities. Radio loud quasars have $\nu F_{\nu}(6 cm)/\nu F_{\nu}(4400 \text{\normalfont\AA}) \geq 10$ and present $\sim 5\%$ of all quasars \cite{banados}.

There are three main parts in the structure of the active galactic nucleus.
The first part is compact central source of continuous radiation and high-energy particles -- accretion disk, which is responsible for all manifestations of powerful nuclear activity. The second and the third components are
shells of ionized gas clouds -- the region of broad spectral lines and the region of narrow spectral lines in a heterogeneous density gas disk \cite{karas}. Broad lines originate in the innermost central region (pc in size), so-called broad line region (BLR), 
while narrow lines are formed in the outer, more extended (100 pc), narrow-line region (NLR).
A typical X-ray spectrum of AGN is composed from the following components \cite{fabian}:
 
- an underlying power-law continuum due to thermally Comptonized soft photons, 

- a soft excess at low energies below 1 keV due to thermal (black-body) emission from an optically-thick accretion disk, 

- a fluorescent/recombination FeK$\alpha$ line, 

- a Compton hump due to X-ray reflection from the disk. 

Emissivity of the disk has an important role in the line and continuum shapes. Although the standard model of the accretion disk does not predict the power-law for the disk emissivity, such law is usually accepted in the case of the hard X-ray emission from the inner parts of AGN accretion disks \cite{fabian89,nandra97,nandra99}. In the case of the disk outer parts (i.e. for optical emitting region), the black-body emissivity law is assumed.

The unified model for AGNs is widely accepted. The physics of black hole, accretion disk, jet, and obscuring torus is convolved with the geometry of the viewing angle and can explain most of the apparent disparate properties of active galaxies \cite{anton, beckmann}. Type I and type II diversity in AGNs emerges as a result of varying orientation relative to the line of sight of similar objects. An observer can view the central engine directly (type I AGN) or through the optically thick torus (type II AGN). Type I objects exhibit the straight physics of AGNs with no absorption and type II objects arise when the line of sight is obscured \cite{hickox, baetriz}. 
A crucial argument that approves the unified scheme is observation of heavily obscured powerful quasars called QSO type II \cite{typeII}. 

In type I AGN, we directly see the region of broad lines and the ultraviolet excess radiation of the accretion disk. The clouds closer to the SMBH have very broad permitted recombination lines (full width at half maximum: $FWHM \geq 3000 km/s$), and also have blue broad-band optical continua \cite{interm}. In addition, X-rays from the corona of hot gas located above the accretion disk is observed in the immediate vicinity of the SMBH. Type I AGNs may have narrow (forbidden) lines in the visible range; intense radiation in the infrared region -- radiation processed in a gas-dust torus coming from the central regions. Type I QSO's have spectra like Seyfert 1 galaxies, but stellar absorption features are not easily detected and narrow lines tend to be weak \cite{unif}.

In active galactic nuclei of type II, the central region is obscured by structures of a gas-dust torus and a galaxy. Ultraviolet and soft X-rays are absorbed. In the visible range, the spectrum of the surrounding galaxy dominates the optical emission of the accretion disk. In the spectra of type II AGNs there are only narrow lines of radiation in the IR region and hard X-rays from the corona. In AGN of type II there is very weak radiation from the accretion disk (ultraviolet excess) in $10^{-3} - 1$ keV range due to absorption by a large column density of gas (in torus) \cite{typeII}, \cite{zakamska} - \cite{alex}. QSOs of type II have been predicted to have narrow permitted lines ($FWHM \sim 500 km/s$); they usually possess red continua, powerful hard X-ray emission and a high equivalent width FeK$\alpha$ line \cite{gisel}.

Kleinmann et al. in 1988 discovered the first type II QSO \cite{kleinmann}. The absence of intense optical radiation in the spectra of type II QSOs greatly complicates their search and detailed study, as well as the study of the entire population of AGNs as a whole. To study such objects, one has to use infrared and X-ray ranges of the electromagnetic spectrum. The successful finding of an optically obscured central AGN in the starburst NGC 6240 \cite{Vignati} made X-rays the obvious wavelength to uncover the Type II QSO phenomenon \cite{typeII}. Some papers discuss the fraction of absorbed AGNs as a function of luminosity and redshift but there is no evidence of such dependence \cite{Dwelly, Tozzi, Akylas}. 

Most of the X-ray sources from deep XMM-Newton 
\footnote{X-ray Multi-Mirror Mission, European Space Agency.}
and other X-ray surveys are absorbed by column densities $10^{21} < nH/cm^{-2} < 10^{23}$, with hard X-ray luminosities of $10^{42} <L_{2-10keV}/(erg/s) < 10^{44}$ \cite{Comastri, Szokoly, Mateos, Mainieri, krumpe}. These values are also found in Compton-thin absorbed ($10^{21} < nH/cm^{-2} < 10^{24}$) Seyfert 2 galaxies at low redshifts \cite{krumpe}.

By analogy with the intermediate types of Seyferts, there should also exist intermediate types of quasars, which display both narrow and broad components to the permitted lines or weak broad $H_{\alpha}$ and $H_{\beta}$ lines, or only weak broad $H_{\alpha}$ detectable.

A prominent feature in the hard X-ray spectra of AGN is a narrow emission line at 6.4 keV FeK$\alpha$ line (there are two components of the line \cite{popov, Fabian}: FeK$\alpha 1$ at 6.404 and FeK$\alpha 2$ at 6.391 keV). The equivalent width (EW) of the line is $\sim 100 - 200$ eV. The FeK$\alpha$ shell is dominated by emission from cold material (Fe I-XVI). In the case of high ionization states of He- and H-like iron ions, the FeK$\alpha$ line energy is increased only to 6.7 and 6.9 keV, respectively \cite{Katsuji}. Presence of both components in spectrum is possible, but it is not possible to separate them using the resolution of $\sim 120 - 150$ eV at 6 keV provided by the CCDs (Charge-Coupled Devices) in the XMM-Newton \cite{Singh}. The line is believed to originate in the inner part of the accretion disk due to fluorescence. The rotation of the accretion disk leads to the Doppler-splitting of the line on the approaching and
receding sides of the disk relative to the observer. The resultant line profile has a blue peak and a bump with an extended red tail, each feature being sensitive to the SMBH mass, spin etc.
The blue peak is very narrow, symmetric and bright (emitting material is rotating slower at larger distances from the black hole), while the red one is wider, asymmetric and much fainter (emitters are located at the lower radii of the disk, i.e. closer to the central black hole)\cite{popov, Singh}. 

Different models of the X-ray absorbing/obscuring regions were developed in order to study how much warm absorbers can change the FeK$_\alpha $ spectral line profile.
Here we will discuss only the most demonstrative model given by P. Jovanovi{\'c} and L. \v C. Popovi{\'c} \cite{Jovanovi}. When the X-ray radiation from approaching side of the disk is significantly absorbed/obscured by the absorbing region, there is a very strong absorption of the iron line. Absorption region is considered to be composed of a number of individual spherical absorbing clouds with the same small radii. The emission FeK$\alpha$ line looks redshifted at $\sim 5$ keV and is followed by strong absorption line at $\sim 7$ keV, which indicates the P Cygni profile of the iron line. Thus, this model can satisfactorily explain the P Cygni profile of the iron line in the case when approaching side of the accretion disk is partially blocked from our view by the X-ray absorbing/obscuring material, while the rest of the disk is less absorbed/obscured and therefore is visible \cite{popov}. An evidence for a P Cygni profile of the FeK$\alpha$ line in Narrow-line Seyfert 1 (NLS1s) galaxies was found by Done et al.  \cite{done}. NLS1 have strong Fe II due to high density region like other BLRs, strong excess below 1--2 keV and rapid X-ray variability.

Observations of Low Ionization Broad Absorption Line (LoBAL) quasars, for example, Mrk 231 \cite{Braito} and H 1413+117 \cite{Chartas} confirmed the presence of X-ray absorbers in these objects. The X-ray emission could be significantly absorbed by an outflowing wind. These flows may originate from a geometrically thin accretion disk, and/or may be associated with the obscuring $\sim 1$ pc torus either by direct X-ray heating or by radiation pressure on trapped infrared radiation \cite{Elitzur, Krolik}. The X-ray spectrum of 1H 0707-495 with the best-fit model which involves a P Cygni profile for the iron features was done by Done et al. \cite{done}.

As seen above, the definition of the types of quasars is still somewhat inaccurate (see also \cite{krumpe, Sihan}). Throughout this paper, we use the type I/II classification based mainly on optical spectra, as it was developed historically. A type II QSO should be an object with narrow forbidden and narrow permitted emission lines in the optical spectrum \cite{krumpe, baldvin}. Objects with less secure type II belong to the tentative identification as sources with ambiguous optical identification, but we give an individual comment for all sources.

\subsection {The Xgal20 catalog} 

\indent \indent The study of extragalactic X-ray sources using the data from XMM-Newton space observatory is one of the main directions of modern X-ray astronomy. XMM-Newton has the largest effective area among all X-ray satellites and can simultaneously obtain images, light curves and spectra of sources in soft and medium X-ray ranges. XMM-Newton is equipped with three European Photon Imaging Cameras (EPIC) CCDs -- two of the cameras are Metal Oxide Semi-conductor (MOS1, MOS2), EPIC-pn (p-n junction) type camera (PN), Reflection Grating Spectrometer (RGS) and EPIC Radiation Monitor (ERM) \cite{streder} - \cite{xmmhand}. 

XMM is used to study different types of AGN, their physical processes and structures, detect variability. We used data from PN camera because pn-type CCDs are mounted in the focal plane therefore the photon beam is directed onto the camera without interruptions. The PN camera has an entire field of view 30 arcmin, effective detector area 1227 $cm^2$, angular resolution -- the FWHM of the point spread function (PSF) is 6.6 arcsec, energy range from 0.2 to 15 keV \cite{Lumb}, but effectively works only in the range up to 10 keV. An energy resolution is $E/dE=20 - 50$. The data extraction was performed with the Science Analysis System (SAS) \cite{sas} version 16.0.0.

In our work we used the latest published version of the XMM catalog, 4XMM-DR9 \cite{webb}. It is based on XMM-Newton EPIC data, which became publicly available on March 1st, 2019. 

A cross-correlation was made between the 4XMM-DR9 catalog and the Hyper-Linked Extragalactic Databases and Archives (HyperLeda) \cite{HyperLeda}. 4XMM-DR9 catalog is modern, large catalog of observations containing 550124 unique sources covering 2.85 \% of the sky, and HyperLeda contains about 1.5 million galaxies. As a result, we obtained a sample of more than 5000 X-ray galaxies, most of which contain low-luminosity active galactic nuclei. Based on this sample, galaxies with an X-ray flux greater than $F \geq 10^{-13}~\mbox{erg}/(\mbox{cm}^2\cdot\mbox{s})$ were selected. Because it is easier to construct an informative spectrum for such sources, they are of particular interest. An identified and classified catalog of 1172 manually verified galaxies, the X-ray Galaxy Catalog Xgal20, was created \cite{zadorozhna, zadorozhna2}. Most galaxies in Xgal20 have an active X-ray nucleus, Seyfert galaxies predominate at short distances among them. However, we were interested in quasars for their more detailed classification by type. Only 11 identified quasars were included in the catalog.

The 4XMM catalog website can be found at http://xmm-catalog.irap.omp.eu, 
the HyperLeda database at http://leda.univ-lyon1.fr.

In our work we also used the NASA/IPAC extragalactic database NED
\footnote{https://ned.ipac.caltech.edu},
the SIMBAD astronomical database -- the Centre de Donn{\'e}es astronomiques de Strasbourg (CDS) \cite{SIMBAD} 
\footnote{http://simbad.u-strasbg.fr}, the Sloan Digital Sky Survey (SDSS-IV) Data Release 16 (DR16) \cite{sdss}
\footnote{https://www.sdss.org/dr16}, the Xgal20 is available at sites.google.com/view/xgal.

\section {The X-ray data and general properties of 11 QSOs sample}   

\indent \indent Our sample  consists of 11 QSOs  observed by  XMM-Newton, and selected from the Xgal20 catalog. The redshifts of the objects range from 0.022 to 1.695 and the Galactic equivalent column density is below $6.3\times10^{20}$~cm$^{-2}$ but in each case it was checked separately. The majority of the sources, 6 out of 10, are classified as Radio Quiet QSOs. The X-ray luminosity covers a range of $10^{41}<L_{2-10keV}<10^{46}$ erg/s.

While the atomic line emission in the optical, UV, and X-ray probe the circumnuclear medium surrounding the SMBH, the shape of the X-ray spectrum ($\sim $ 2--10 keV) is described by $\Gamma$ -- photon index. $\Gamma$ is a critical parameter to constrain competing models for the mechanisms of X-ray continuum emission \cite{Dewangan}. 

The earlier studies suggested that the X-ray emission of quasars is well described by a power-law with photon index $\Gamma \sim 1.5$ for radio-loud quasars and $\sim 2.0$ for radio quiet quasars. Large samples of AGNs have been studied by Walter, Fink \cite{Walter}, Wang, Brinkmann, Bergeron  \cite{Wang}, and by Rush et al. \cite{rush} using the ROSAT PSPC (R{\"o}ntgensatellit Position Sensitive Proportional Counters) \cite{Dewangan}.

The continua of most QSOs are satisfactorily fitted by a power law with low energy absorption. The mean value of the hard X-ray photon index of the power law in our sample is $2.17\pm 0.07$ in agreement with previous results. $\Gamma$ is independent of 2--10 keV luminosity; however, a large scattering around the mean value is clearly detected. We checked for the presence of other spectral features both in the hard and soft XMM bands. The presence of soft excess is almost ubiquitous but a separate emission model for its fitting was needed in some cases. It was fitted with a {\it black body} model for 3 objects. A {\it multi-temperature blackbody disk} emission, a combination of 2 {\it black bodies} or {\it bremsstrahlung} popular models were checked but they don't fit the soft excess for any QSO in our sample (for X-ray fitting models and their parameters see also \cite{Arnaud}). 

For 5 objects, the presence of a complex absorption/emission pattern in the soft X-ray band requires a multi-component ionized gas models. An emission spectrum from collisionally-ionized diffuse gas calculated using the {\it apec} or {\it mekal} model. The models associate soft X-ray emission with emission lines of ionized elements within the
thin, highly ionized NLR. These are irradiated by the original continuum from the central region, leading to the formation of many emission lines. Each model includes line emissions from several elements (C, N, O, Ne, Mg, Al, Si, S, Ar, Ca, Fe, Ni). 

Two QSOs show evidence of a FeK$\alpha$ emission line, detected lines present a high equivalent width. We fit FeK$\alpha$ line with redshifted Gaussian profile {\it zgauss} model. No significant trend in the line energy as a function of the 2--10 keV luminosity has been found.
   
\subsection {Observations and spectroscopy} 

\indent \indent $\bf{3C 189}$ (NGC 2484, B2 0755+37) is a radio galaxy with jet, class Fanaroff-Riley type I radio source. The radio structure was shown in de Ruiter et al. \cite{deRuiter} to be that of a weak, one-sided, jet within a diffuse, low surface-brightness, envelope. But sensitive, well-resolved Very Large Array imaging shows that jet in the source has a two-component structure transverse to it axis. Close to the jet axis, a centrally-darkened counter-jet lies opposite to a centrally-brightened jet, but both are surrounded by broader collimated emission that is brighter on the counter-jet side \cite{Laing}. The ROSAT High Resolution Imager (HRI, $\sim 0.2 - 2$ keV) data are consistent with an unresolved source of luminosity $L\sim (3.7\pm0.4)\cdot 10^{42} erg~s^{-1}$ \cite{Canosa, Worrall}. NGC 2484 alternative classification in NED is Low-Excitation Radio Galaxy (LERG). 
Optical luminosity of this object was estimated by the Fermi Large Area Telescope: $M_{V}=-27.0$.

In the SDSS catalog this object's number is SDSS J075828.10 + 374711.8.
Its optical spectrum was automatically obtained by the SDSS spectrograph, which lead to evaluation its class -- galaxy broadline and redshift z=0.04.
Broad absorption lines Na, Mg, Ca, K can be clearly seen, as the same as absorption hydrogen (H$\alpha$ line and Balmer lines H$\beta$, H$\gamma$, H$\delta$, H$\epsilon$, H$\zeta$), weak emission hydrogen lines. One can also see the narrow forbidden lines of oxygen, neon, nitrogen, sulfur, argon. SDSS J075828.10 + 374711.8 seems to have strong absorption features. It is not typical for QSOs with broad absorption lines \cite{Turnshek, zheng}, that this object is an intense radio source.

\begin{figure}[!htb]
	\minipage{0.5\textwidth}
	\includegraphics[width=\linewidth]{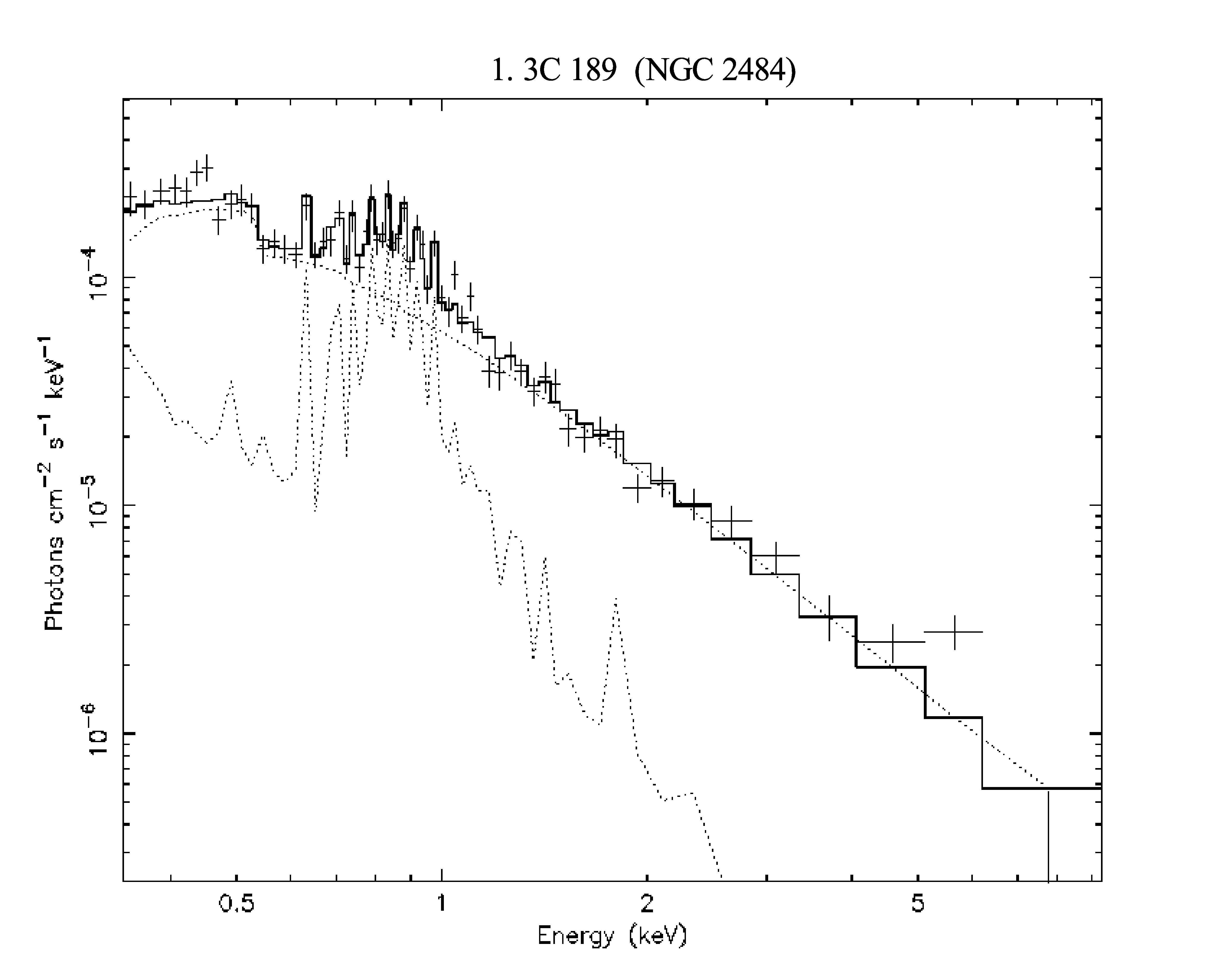}
	\endminipage
	\hfill
	\minipage{0.5\textwidth}
	\includegraphics[width=\linewidth]{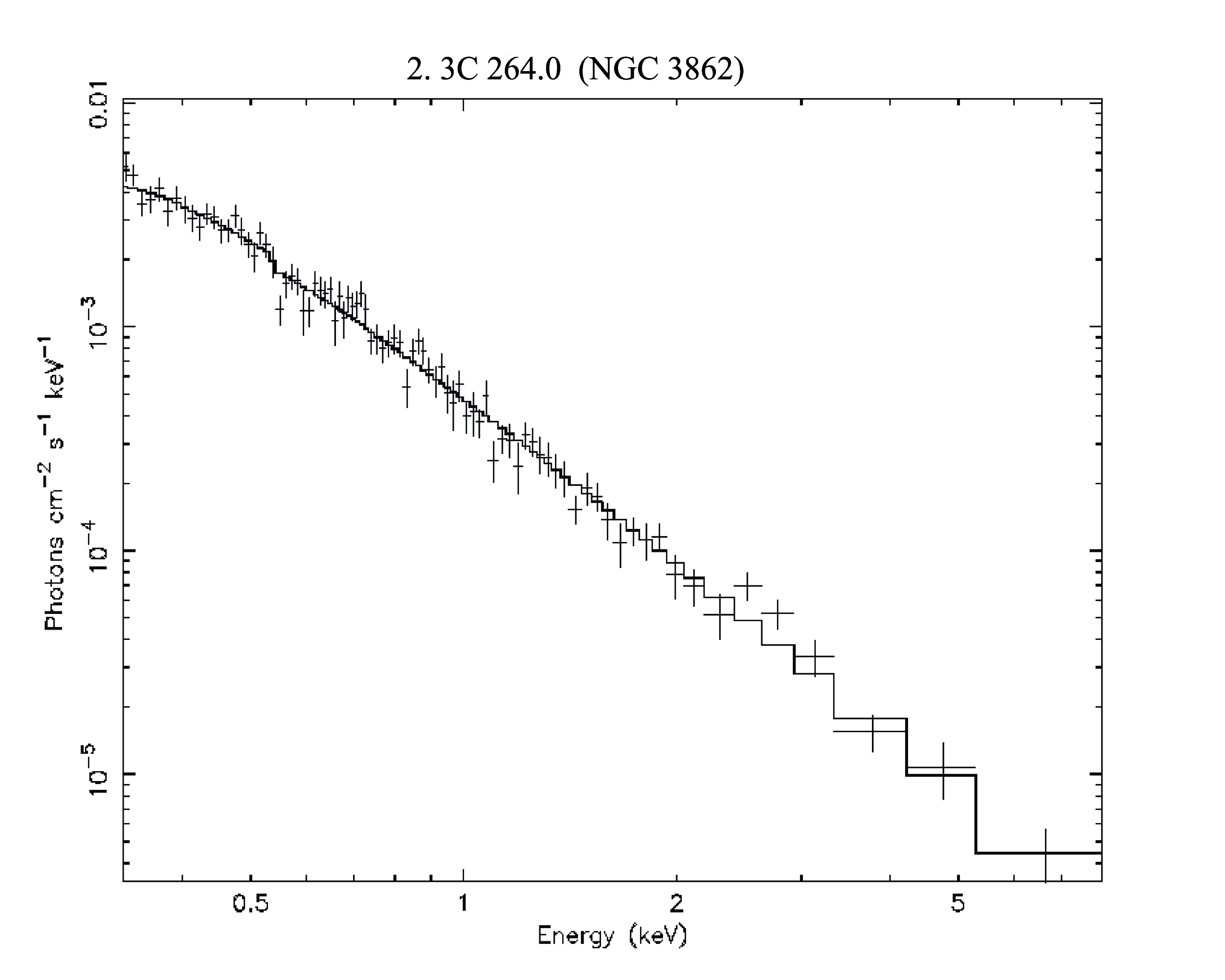}
	\endminipage
	\hfill
	\minipage{0.5\textwidth}
	\includegraphics[width=\linewidth]{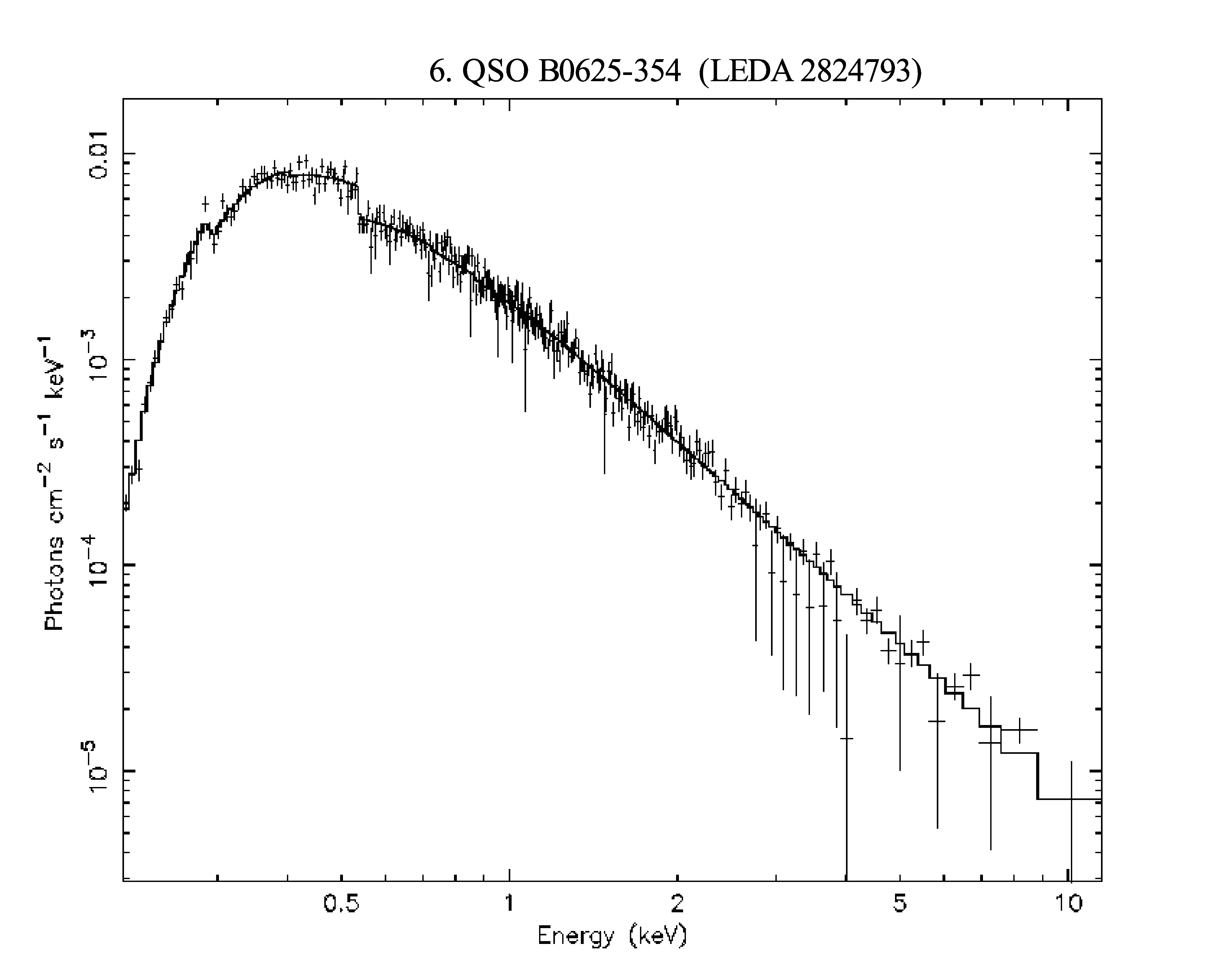}
	\endminipage
	\hfill
	\minipage{0.5\textwidth}
	\includegraphics[width=\linewidth]{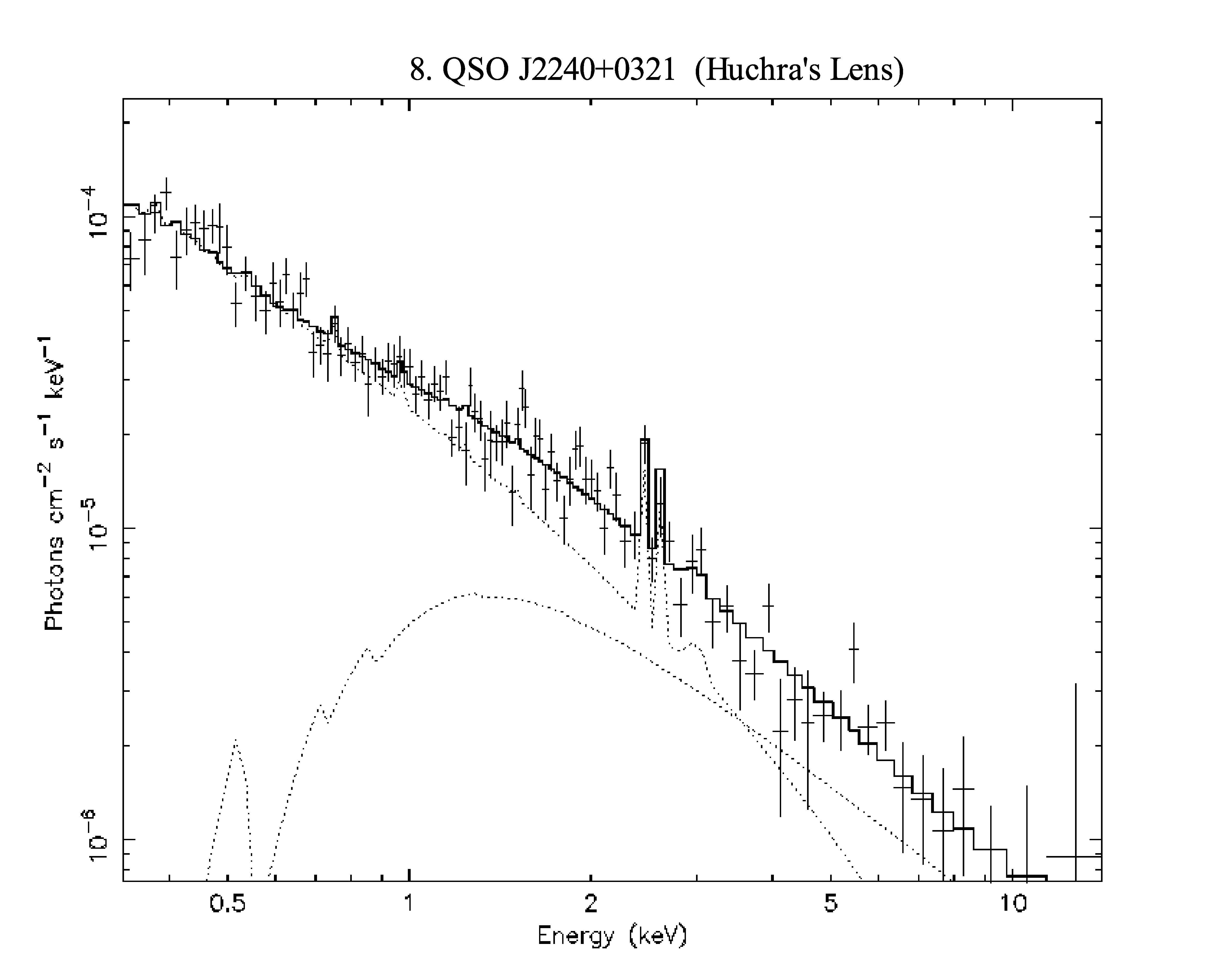}
	\endminipage
	\hfill
	\minipage{0.5\textwidth}
	\includegraphics[width=\linewidth]{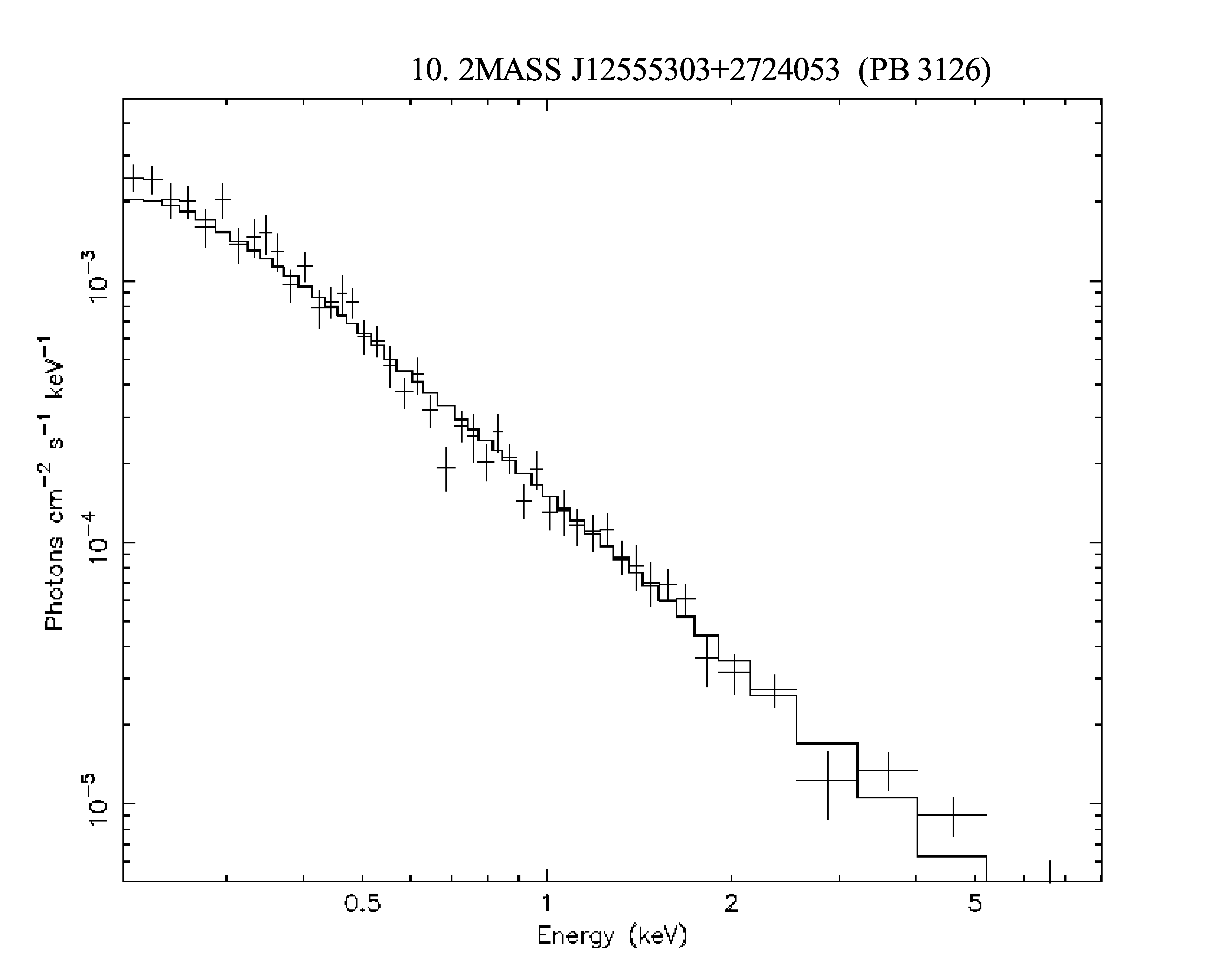}
	\endminipage
	\caption{The XMM-Newton EPIC PN spectrum of the 5 type-I QSOs from Xgal20 sample in the 0.1-10 keV band plotting with the best-fit models.}\label{s1}
\end{figure}

3C 189 was observed by XMM Newton one time (see Table \ref{Sample Data}, Table \ref{master}), the observation was fitted with {\it powerlaw*phabs + mekal} model (see Pic. \ref{s1}, Table \ref{Spectr}) with good statistical probability 64.43/58 and null hypothesis probability $26 \% $. Soft X-ray edges due to ionized absorption/emission, has been observed at $\sim 0.6 - 1$ keV.

3C 189 is a representative of radio-loud type I quasar.

$\bf{3C 264.0}$ (NGC 3862, LEDA 36606) is among the closest known bright radio galaxies with an optical jet (also detected at X-ray energies by Chandra X-ray Observatory) at a redshift of 0.0217 and hence had distance of 94 Mpc \cite{Perlman}. It is hosted by NGC 3862, a large elliptical galaxy offset to the south-east from the center of the cluster Abell 1367. On large scales it exhibits a twin-tail radio structure extending to the north-east \cite{Bridle, Perlman}. On arcsec scales it consists of a compact core, with a one-sided, nearly knot-free jet extending also to the north-east and a weak counter jet \cite{Lara, Perlman}. The optical jet counterpart extends for only roughly 2 arcsec beyond the galaxy core. 3C 264.0 is classified as an Fanaroff-Riley type I radio source, it's alternative classifications in NED are Low-Excitation Radio Galaxy (LERG) and Low Ionization Nuclear Emission Line Region (LINER). It is the member of the Third Catalog of active galactic nuclei detected by the Fermi Large Area Telescope \cite{Fermi} with optical absolute magnitude $M_{V}=-26.2$.

3C 264.0 has a number SDSS J114505.01+193622.8. It was automatically classified by the SDSS spectrograph as 'galaxy AGN broadline'.  The spectrum is very similar to the previous quasar with the same identified lines -- broad absorption Na, Mg, Ca and K, absorption hydrogen (H$\alpha$, H$\beta$, H$\gamma$, H$\delta$, H$\epsilon$, H$\zeta$), the narrow forbidden lines of oxygen, neon, sulfur, nitrogen, argon.

This quasar was observed by XMM Newton 3 times (see Table \ref{Sample Data}, Table \ref{master}). The observation ID 0061740101 was selected for processing. Its spectrum was fitted with {\it powerlaw*phabs} model (see Pic. \ref{s1}, Table \ref{Spectr}). The fitting has very good statistical probability 86.02/83 and null hypothesis probability $39 \%$. The X-ray image clearly shows the glow of the cluster's Abell 1367 halo.

So 3C 264.0 is classified as radio-loud type I quasar.

$\bf{3C 338.0}$ (NGC 6166) is an elliptical (cD2 pec) galaxy in the Abell 2199 cluster. It lies 490 million light years away in the constellation Hercules. It has a supermassive black hole at its center with a mass of nearly $30 \cdot 10^9M_{\odot}$ based on dynamical modeling \cite{Magorrian}. 
NGC 6166 is classified as an Fanaroff-Riley type I source, which powers two symmetric parsec-scale radio jets and radio lobes.
Jets ans lobes are caused by the infall of gas into its center according to a cooling with the rate of 
$200 M_{\odot}$ of gas per year \cite{Matteo}. 
Such accretion rate corresponds to nuclear X-ray luminosity of $L\sim 10^{44} erg/s$ -- low-luminosity. NGC 6166 is the member of the Third Reference Catalog of Bright Galaxies \cite{Third} with absolute magnitude $M_{V}=-23.63\pm0.40$.

NGC 6166 is a radio source with Ultra Steep Spectrum (USS; $\alpha \geq 1.30$, $S \sim \nu^{\alpha}$, $S_{1400}>10 mJy$)\cite{DeBreuck}. 
It's alternative classifications in NED are Low-Excitation Galaxy (LEG), Narrow-Line Radio Galaxy (NLRG). In unified models NLRG is the radio loud equivalent of type II QSO. 

3C 338.0 with number SDSS J162838.24+393304.5 has the interactive spectrum obtained by the BOSS spectrograph, class 'galaxy AGN broadline' and z=0.031. All the same lines are recognized in the spectrum, as for previous QSO's, but their shape and width are different. The spectrum of this quasar differs from the previous two, just as the spectra of Seyfert 1.5 would differ from Seyfert 1. The absorption lines Na, Mg, Ca, K, emission hydrogen Balmer lines H$\beta$, H$\gamma$ are narrow. Also narrow emission forbidden components are easily seen.

\begin{figure}[!htb]
    \minipage{0.5\textwidth}
	\includegraphics[width=\linewidth]{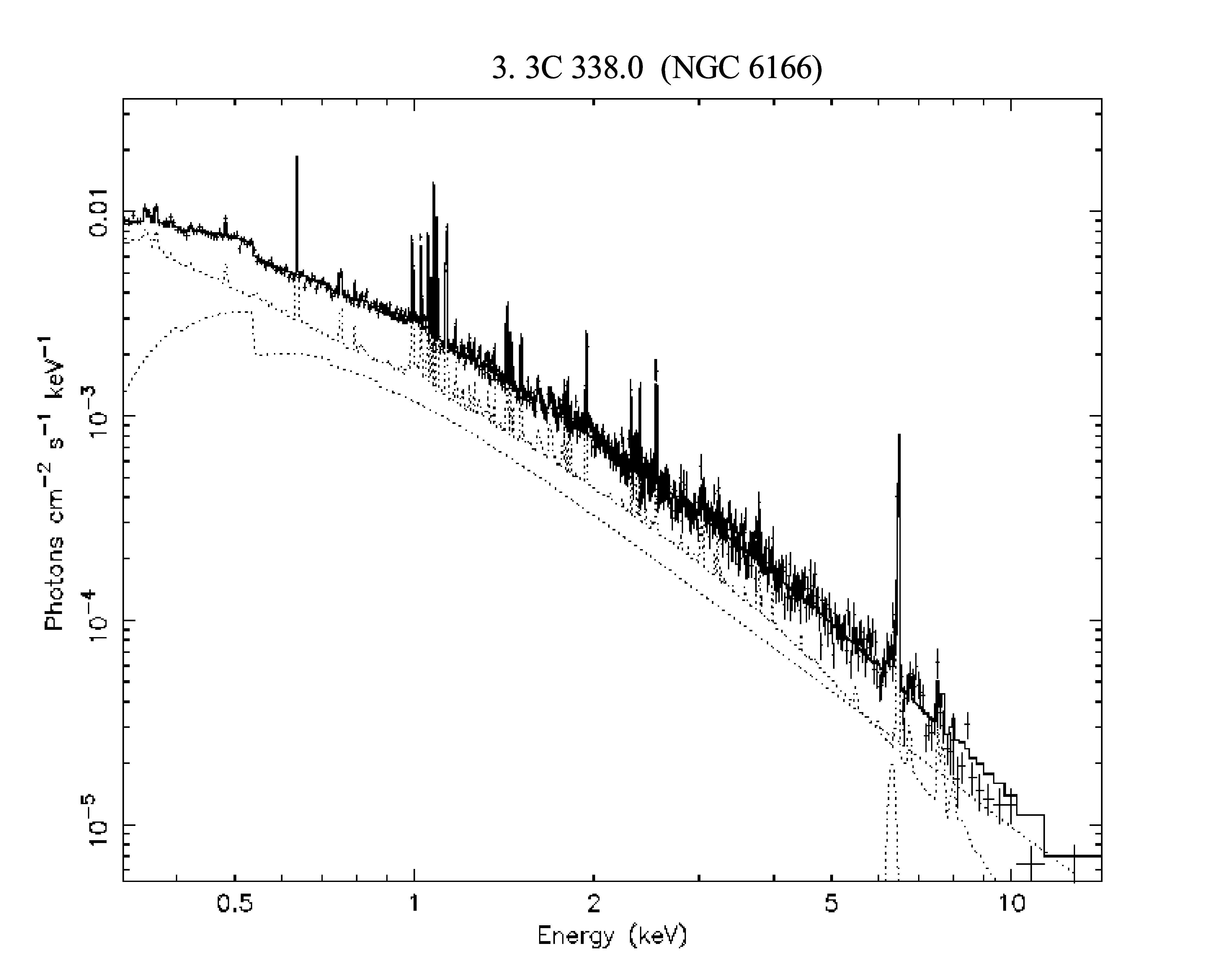}
     \endminipage
     \hfill
     \minipage{0.5\textwidth}
     \includegraphics[width=\linewidth]{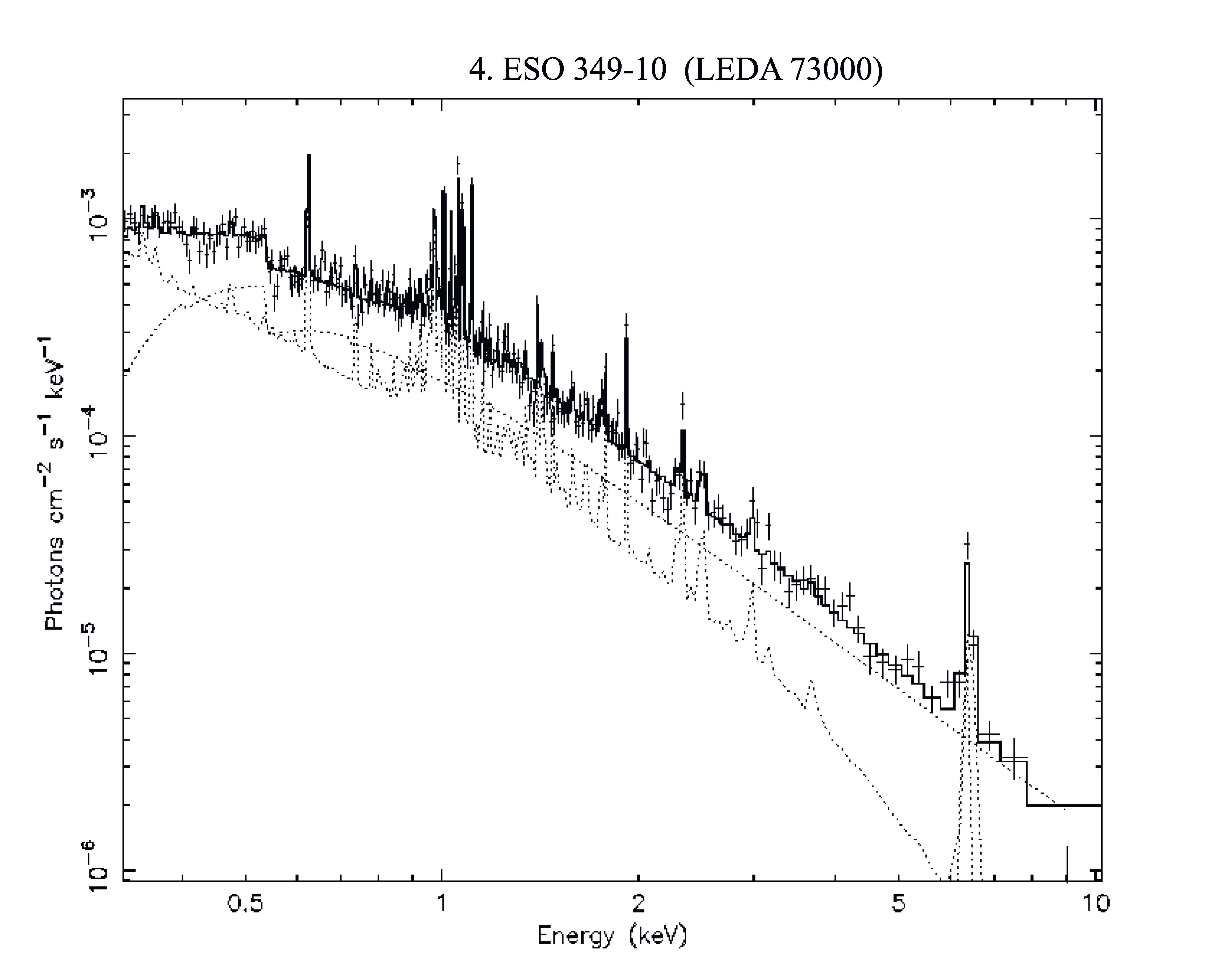}
     \endminipage
     \hfill
     \minipage{0.5\textwidth}
     \includegraphics[width=\linewidth]{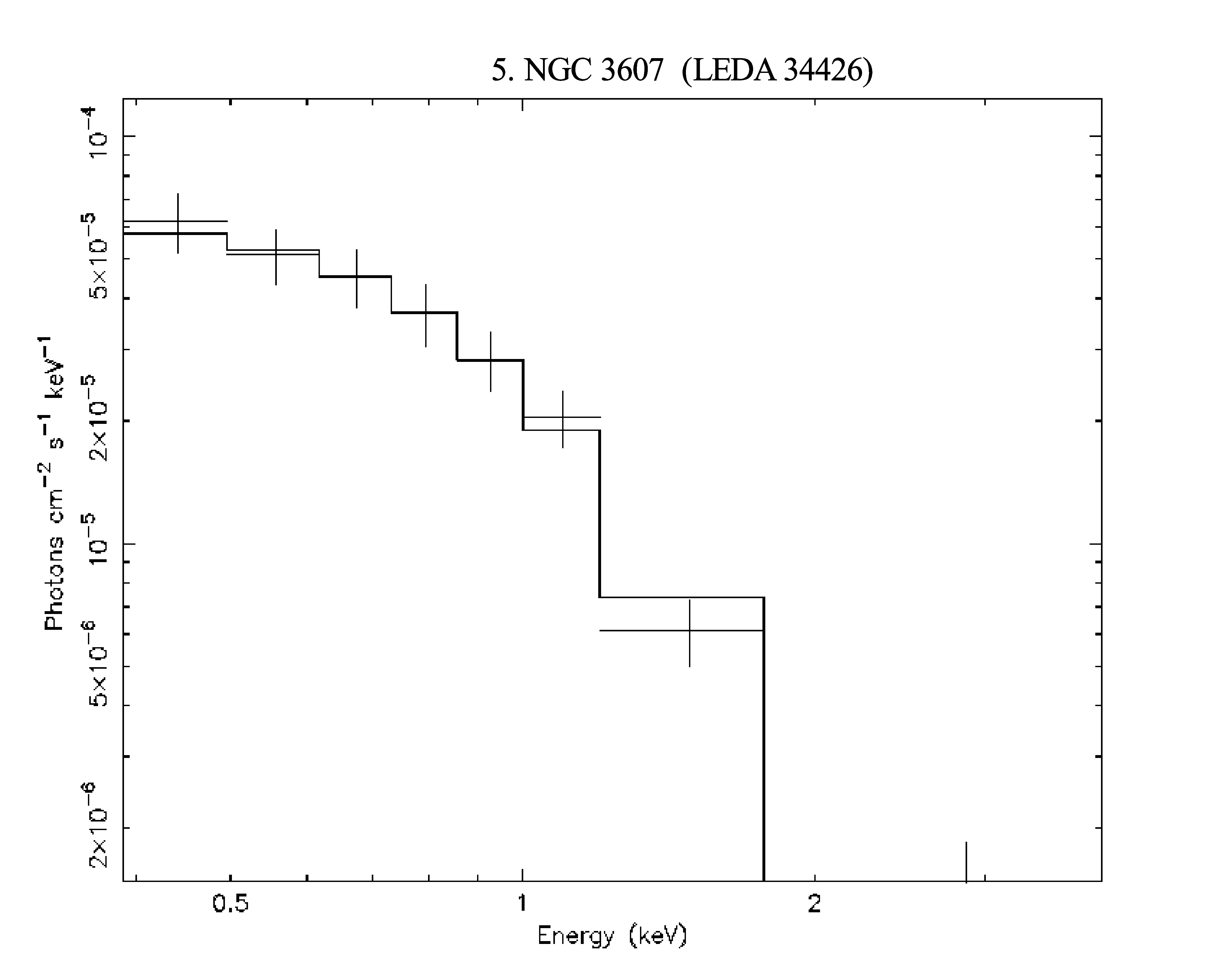}
     \endminipage
     \hfill
     \minipage{0.5\textwidth}
     \includegraphics[width=\linewidth]{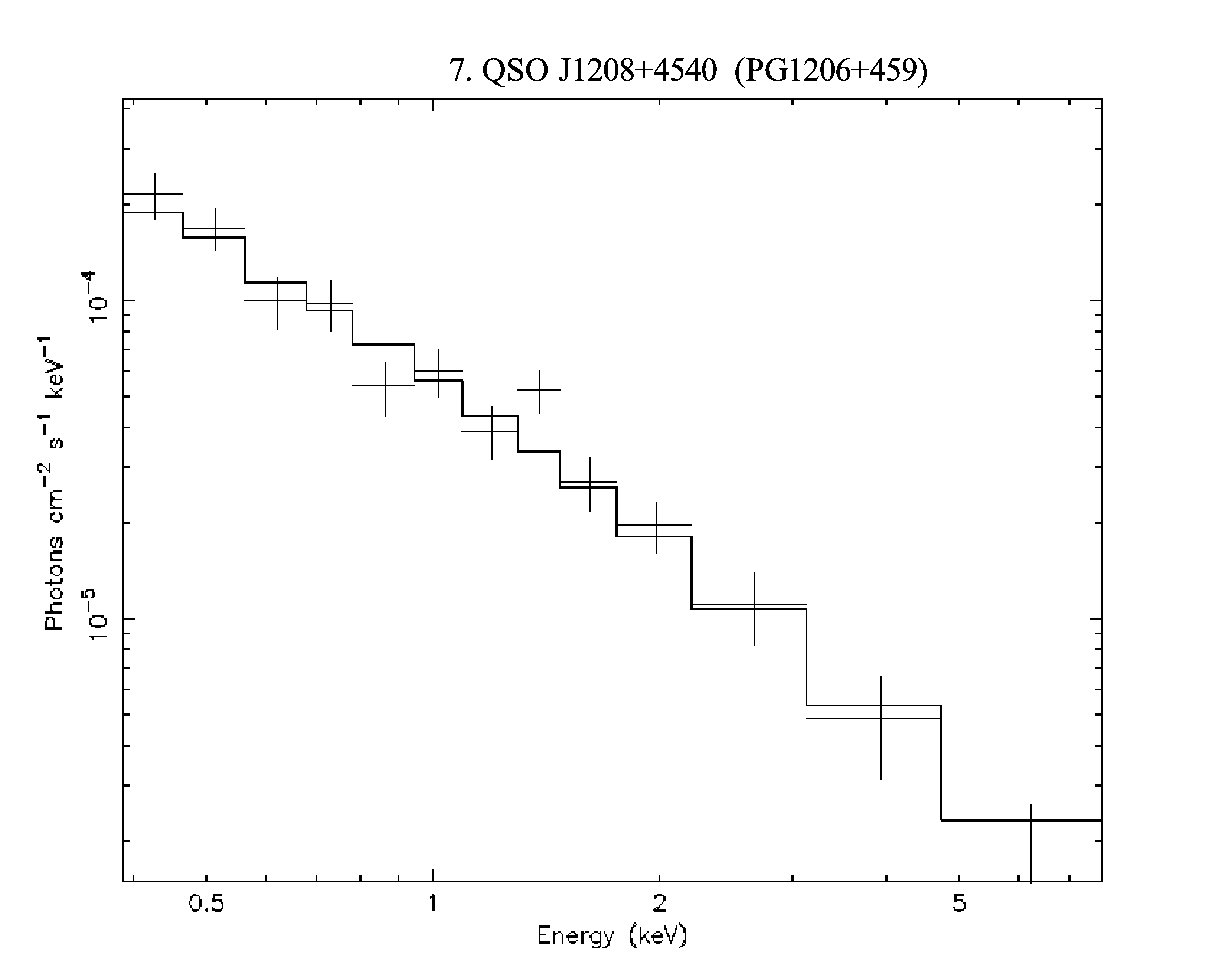}
     \endminipage
     \hfill
     	\minipage{0.5\textwidth}
     \includegraphics[width=\linewidth]{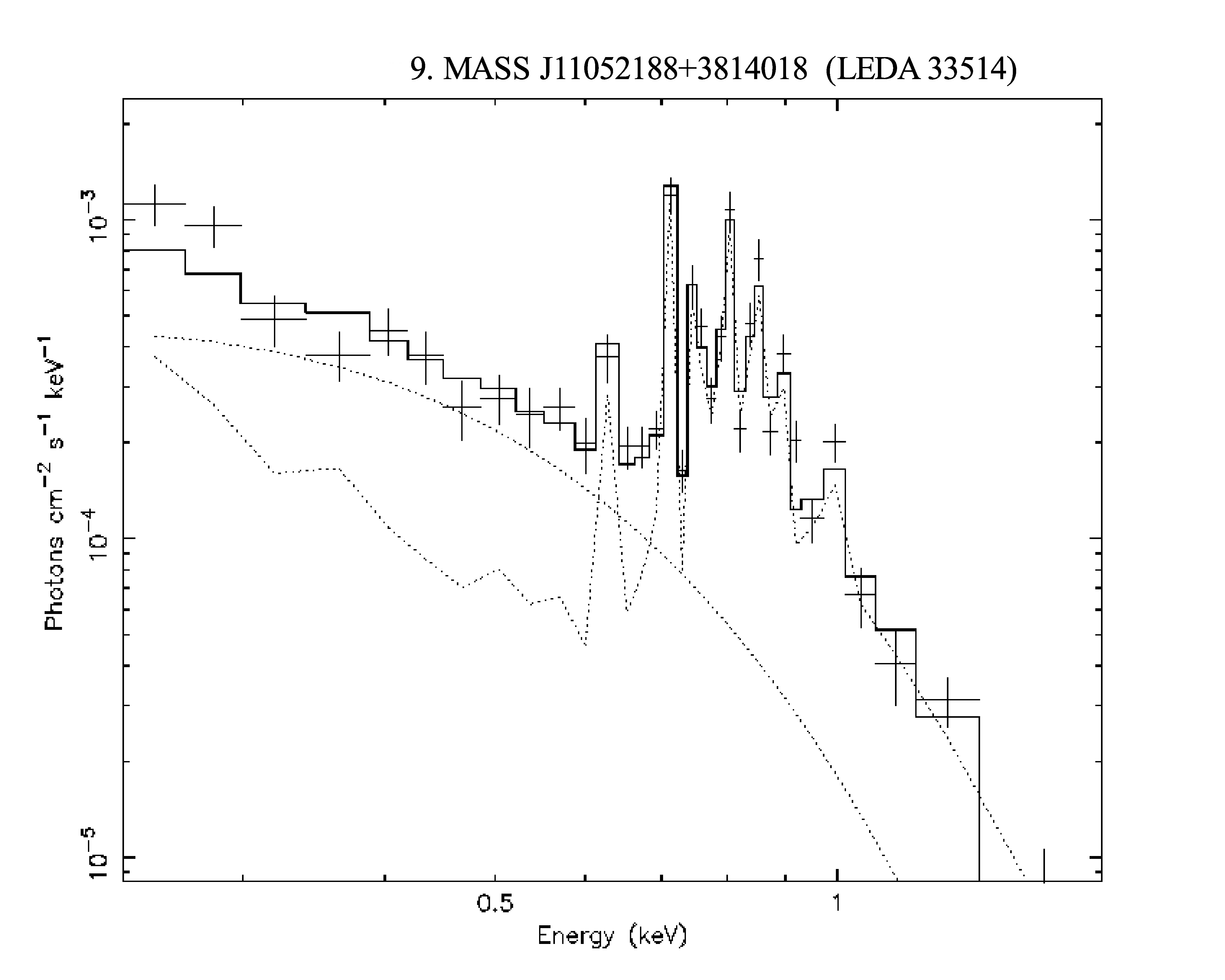}
     \endminipage
     \hfill
     \caption{The XMM-Newton EPIC PN spectrum of the 5 type-II QSOs (obscured) from Xgal20 sample in the 0.1-10 keV band plotting with the best-fit models.}\label{s2}
 \end{figure}

It was observed by XMM Newton 5 times (see Table \ref{Sample Data}, Table \ref{master}). For our processing the observation with ID 0723801101 was selected, its spectrum was fitted with complex model (see Pic. \ref{s2}, Table \ref{Spectr}), {\it powerlaw*phabs+apec+zgauss}. The model has statistical probability 961.27/740 and bad null hypothesis probability $(\sim 10^{-7})$ due to the large number of fit points. 3C 338.0 shows a complex spectrum, dominated by ($nH\sim 1.2 \cdot10^{21}$~cm$^{2}$) ionized obscuration. Single soft X-ray edge is caused, most likely, by ionized oxygen absorption.
The edge has been observed at $\sim 0.6$ keV, probably originating in the same warm medium responsible for the absorption in the soft band. The narrow emission FeK$\alpha$ line has been observed at $\sim 6.5$ keV, it suggests an origin from cold (FeI-XV) matter distant from the inner disk region.  

With some caution, we would classify 3C 338.0 as radio-loud and an intermediate class between type I and type II (rather type II). The quasar seems to be obscured, although this requires additional consideration.

$\bf{ESO 349-10}$ (LEDA 73000, PKS 2354-35) is the elliptical galaxy (E-S0, cD4) identified with a Fanaroff-Riley type I radio source at a redshift of z = 0.04905 (distance $\sim 222$ Mpc, $M_R=-24.53\pm0.51 $). It contains an optical LINER-type nucleus. ESO 349-10 is a member in a group Abell 4059 and classified by NED as brightest cluster galaxy (BrClG) with optical line emission.

It is an Ultra Steep Spectrum radio source. Singh, V., Beelen, A., et al. \cite{Singh2} show that the criterion of ultra steep spectral index remains a method to select populations of weaker radio-loud AGNs potentially hosted in obscured environments.

This quasar was observed with Spitzer Space Telescope Infrared Array Camera (IRAC) and Multiple Imaging Photometer for Spitzer (MIPS) and appears as a compact nuclear source embedded within diffuse host emission. The Spitzer images have been analyzed in A. C. Quillen \cite{Quillen}, who find fluxes for the BrClG in Abell 4059. He got $L_{IR}< 0.3\cdot 10^{44} erg/s$. Abell 4059 exhibits a similar spectrum with [NII] lines but no H$\alpha$ line as seen from the 6dFGS archive (Six-degree Field Galaxy Survey) \cite{Jones2004, Jones2006}. There is no any evidence of infrared excess from this galaxy. 

We did not find the object ESO 349-10 or Abell 4059 in the SDSS database.

It was observed by XMM Newton 4 times (see Table \ref{Sample Data}, Table \ref{master}). The observation 0723801001 was selected for our analysis. The spectrum was fitted with combined model (see Pic. \ref{s2}, Table \ref{Spectr}), {\it powerlaw*phabs+apec+zgauss}. The fit has good statistical probability 351.16/296 and null hypothesis probability $1.5\% $. ESO 349-10 shows a complex spectrum, dominated by ($nH\sim 1.2 \cdot 10^{21}$~cm$^{2}$) ionized obscuration. Soft X-ray edge, likely due to OVII absorption, has been observed at $\sim 0.6-0.7$ keV. The extrapolation of the power law to energies lower than 2 keV clearly revealed the presence of large deviations due to complex absorption/emission.  
The emission FeK$\alpha$ line has a double structure and has been observed at $\sim 6.7 - 6.8$ keV, it suggests high ionization states iron ions from matter close to the inner accretion disk region. 

ESO 349-10 shows obscuration properties, is classified as radio-loud type II quasar.

$\bf{NGC 3607}$ (LEDA 34426, XBSJ111654.8+180304) is an unbarred spiral galaxy (SA0$^0$) at z= 0.00314, that forms part of the Leo II Group (NGC 3607 Group) of galaxies. Its central black hole has a mass of about $(1.2 \pm 0.4)\cdot10^8 M_{\odot}$ \cite{Gultekin}. NGC 3607 shows narrow emission lines in near-IR and optical spectrum suggestive of star-formation. NGC 3607 has an IR spectrum qualitatively similar to those of LINER 2-type nucleus \cite{ho}. XBSJ111654.8+180304 in SIMBAD is classified as quasar, its alternative classification, for example, in NED is Seyfert 2 galaxy. Absolute magnitude in z band is $-21.9$, which is generally inconsistent with the definition of a quasar. But this object is brighter in near-IR -- M=$-24.28\pm0.31$ (see in NED or 2MASS Large Galaxy Atlas). 

NGC 3607 rather can be defined as elusive AGN \cite{Caccianiga}. A. Caccianiga et al. notice that NGC 3607 X-ray luminosity $L_{2-10 keV}<10^{41} $ erg/s is the lowest among the extragalactic XBS (The XMM-Newton Bright Serendipitous Survey) sources. The "recognition problem" for AGN is critical in the low-luminosity regime, this problem affects mostly absorbed AGN $\sim 40\% $ of type 2 AGN in the survey are elusive, but also a significant fraction of unabsorbed AGN $\sim 8\% $ \cite{Caccianiga}.

Comastri et al. have suggested that an heavily absorbed AGN (Compton-thick with $nH > 10^{24} cm^{-2}$) could be hidden within some XBONG (X-ray Bright Optically Normal Galaxies) \cite{Comastri1, Comastri2}. 

A. Caccianiga et al. also note that the detection of a Compton-thick AGN, using the 2-10 keV energy band may be missed, in particular for local AGN, since the energy cut-off is expected to fall outside the observed interval. The indicators that can be used to assess this AGN type are: the detection of a prominent FeK$\alpha$ emission line or/and a very low flux ratio $F_X/F_{[OIII]}<1$. The latter indicator, called Compton-thickness parameter A. Caccianiga applied to XBSJ111654.8+180304 and not supported the Compton-thick hypothesis for it \cite{Caccianiga}.

The presence of a Compton-thick AGN should produce strong narrow emission lines in the optical spectrum of a source with large X-ray flux (larger than $10^{-13} erg~s^{-1} cm^{-2}$). However, there is not enough evidence of such classification for this object.

NGC 3607 has number J111654.64+180306.3 in SDSS, but no spectrum. 

It was observed by XMM Newton 2 times (see Table \ref{Sample Data}, Table \ref{master}), for our processing observation ID 0693300101 was selected.
The spectrum is very poor, it is not detected  
at energy $> 2$ keV. 
The spectrum was fitted with simply model, which is used to fit the soft excess (see Pic. \ref{s2}, Table \ref{Spectr}), {\it bbody}. The fit has statistical probability 10.6/6 and null hypothesis probability $10 \% $. 

We need additional high-resolution X-ray data to detect or rule out the presence of the FeK$\alpha$ emission line for a more detailed discussion of its Compton-thickness possibility.

We decided to classify NGC 3607 preliminary as radio-quiet type II AGN.

$\bf{QSO B0625-354}$ (LEDA 2824793, PKS 0625-35) is the elliptical galaxy at z= 0.05459 ($M_V=-27.5 $) and a well-known source of gamma radiation that was included in the Online Gamma-Ray Catalog TeVCat with name TeV J0626-354. The discovery of gamma-ray emission from this source was announced by Dyrda et al. at the 34th International Cosmic Ray Conference (ICRC 2015) \cite{Dyrda}. The classification of this object in TeVCat has been BL Lac of unknown type and later it was updated to an AGN of unknown type following the discussion by Rieger and Levinson \cite{Rieger}. 
It's alternative classification in NED is Seyfert 3 (LINER type 1) and Elliptical Weak-line radio galaxy (WLRG) \cite{Lewis}. 

PKS 0625-354 has a black hole of mass approx. $10^9 M_{\odot}$ that is probably accreting in an inefficient mode\cite{mingo}. The source is known as a low excitation line radio-loud AGN, but being a transitional Fanaroff-Riley type I/BL Lac object. 

Object information from The High Energy Stereoscopic System (H.E.S.S.) Colla-boration \cite{HESS} is the following: 

- it is a blazar or moderately misaligned blazar; 

- it has larger value of the redshift when compared to those of the other Very High Energy (VHE) radio galaxies; 

- the intensity of the O[III] emission line has been used to choose between BL Lacs and FR I radio galaxies. O[III] luminosity places this object to the BL Lac category. 

- Significant Doppler boosting is required to reproduce the measured broadband spectral energy distributions (SEDs) with a maximal angle between the jet and the line of sight of $\sim 15 $ deg. The source is well described with a power-law TeV spectrum with spectral index: $2.84 \pm 0.50$.

QSO B0625-354 has no spectrum in SDSS. 

This galaxy was observed by XMM Newton 3 times (see Table \ref{Sample Data}, Table \ref{master}), for processing observation ID 0302440601 was selected. Its spectrum was fitted with simple model, which in most cases describes type I quasars well (see Pic. \ref{s1}, Table \ref{Spectr}) -- {\it powerlaw*phabs}.
Our model has good statistical probability 367.82/308 with null hypothesis probability $1.08\% $. Best fit parameters of the model are $\Gamma = 2.5 \pm 2.3 \cdot 10^{-2}$, $nH\sim 8.3 \cdot 10^{20}$~cm$^{2}$). Soft X-ray edges, likely due to ionized absorption lines, has been observed at $\sim 0.3$ and $\sim 0.7$ keV. 

We see that this AGN was classified with all main types by different authors. 
Our decision is to classify QSO B0625-354 as radio-loud type I quasar.

$\bf{QSO J1208+4540}$ (PG1206+459) is classified as QSO (activity type) and as Hyperluminous infrared galaxy (HyLIRG) \cite{Rowan} at z= 1.16494 ($M_i=-29.26\pm0.50 $).

M. Rowan-Robinson argued that hyperluminous galaxy with emission peaking at rest frame wavelengths $\geq 50\mu$m are undergoing star formation at rates $>10^3 M_{\odot} yr^{-1}$. 
Infrared bump in spectrum corresponds to star formation or/and thermal emission caused by dusty torus. 
But in the same work he showed that for PG1206+459 the Infrared Astronomical Satellite (IRAS) $12 - 16~\mu$m data and the Infrared Space Observatory (ISO) $12 - 200~\mu$m data \cite{haas} are well-fitted by the extended AGN dust torus model and there is no evidence for a starburst \cite{Rowan}. In near-IR it was observed by WISE \cite{wise}. 

In the SDSS catalog it is the object with number SDSS J120858.00+454035.4, good spectrum obtained by the SDSS spectrograph and QSO broadline class. It has typical spectrum for QSO type I. Broad emission of Mg II, Balmer continuum (without Balmer lines) -- ``Small blue bump``, C II], C III] (semi-forbidden lines) can be clearly seen. 

This galaxy was observed by XMM Newton only once (see Table \ref{Sample Data}, Table \ref{master}), observation ID 0302440601. It's spectrum is poor, it was fitted with simple model, which describes type I quasars well (see Pic. \ref{s1}, Table \ref{Spectr}), {\it powerlaw*phabs}. X-ray spectrum parameter of this object is $\Gamma = 1.84 \pm 0.23$. The fit has statistical probability 11.70/10 and null hypothesis probability $31\% $.

QSO J1208+4540 is typical representative of radio-quiet type I QSOs.

$\bf{QSO J2240+0321}$ (Huchra's lens, The Einstein Cross) was discovered by Huchra et al. in 1985 \cite{huchra}, which is one of the most famous gravitational lens systems with a radio-quiet quasar at z = 1.695 that is quadruply lensed by a z$_g$ = 0.039 Sab galaxy. Absolute magnitude in visible spectral region is $-31.42\pm0.54$. 

From the European Southern Observatory (ESO) archive, L. Braibant et al. were retrieved near-infrared spectra of QSO2237+0305 that was obtained in October 2005 with the integral field spectrograph SINFONI (Spectrograph for INtegral Field Observations in the Near Infrared) mounted on the Yepun telescope (UT4) of the Very Large Telescope (VLT). 
Observations were performed using the $3'' \times 3 ''$ field-of-view (FOV) with $0.1 ''$  spatial resolution and the H-band grism, whose spectral coverage goes from 1.45 to 1.85 $\mu m$ and thus includes the H$\alpha$ broad emission line and other high- and low-ionization broad emission lines \cite{Braibant}.

Visible spectra of QSO2237+0305 were acquired with the FOcal Reducer and low dispersion Spectrograph (FORS1) mounted on Unit Telescope N2 of the VLT under program ID 076.B-0197 (PI: Courbin). Visible spectra contains the CIV, CIII] and MgII broad emission lines \cite{Braibant}.

QSO J2240+0321 has number SDSS J224030.22+032130.2 but no spectrum in SDSS.

It was observed by XMM Newton 3 times (see Table \ref{Sample Data}, Table \ref{master}), for processing observation ID 0823730101 was selected. Its spectrum was fitted with complex model (see Pic. \ref{s1}, Table \ref{Spectr}), {\it powerlaw*phabs+mekal+bbody}. The model has good statistical probability 119.05/101 and null hypothesis probability $10.6\% $. Soft excess and some irregularity in the soft X-ray spectrum were fitted with models {\it bbody} and {\it mekal}.

According to all the above, we conclude that QSO J2240+0321 is an exceptional representative of radio-quiet type I QSOs.

$\bf{2MASS J11052188+3814018}$ (LEDA 33514) is located at the edge of the field of view near the bright blazar Markarian 421. Quasar LEDA 33514 itself is clearly visible in optics (R=11.3, M=-24.3). It is the core of a spiral galaxy and in X-rays looks like a fairly clear spot. This quasar is close, z= 0.02847, radio-quiet, type-II AGN (as classified in NED) \cite{Veron}. 

2MASS J11052188+3814018 with number SDSS J110521.89+381401.7 has the interactive spectrum obtained by the SDSS spectrograph and classified by it as  starforming galaxy. The spectrum is very similar to the 3C 338.0 in this range. 
It has the same lines: narrow absorption lines Na, Mg, Ca, K, emission hydrogen Balmer lines H$\beta$, H$\gamma$, emission forbidden oxygen, neon, nitrogen, sulfur, argon are easily seen. It shows only narrow lines in spectrum.

This region was observed by XMM Newton 131 times (because of the nearby bright blazar) (see Table \ref{Sample Data}, Table \ref{master}), for processing observation ID 0153950701 was selected, its spectrum was fitted with complex model (see Pic. \ref{s2}, Table \ref{Spectr}), {\it apec+bbody}. It has statistical probability 40.50/30 and null hypothesis probability $9.6 \%$. The spectrum is absent at energy $< 2$ keV. Some soft X-ray edges, most likely, ionized absorption/emission, has been observed at $\sim 0.6 - 1$ keV, which shown some ionized obscuration.  

We classify 2MASS J11052188+3814018 as radio-quiet type II QSO.

{\bf 2MASS J12555303+2724053} (PB 3126) is a member of the Coma Cluster (Abell 1656), QSO at z = 0.31605 ($M_z=-23.95 \pm 0.50$), alternative classification is Seyfert 1 \cite{lin}. It is included in the sample LoTSS/HETDEX (Hobby-Eberly Telescope Dark Energy Experiment) of quasars selected by their optical spectra in conjunction with sensitive and high-resolution low-frequency radio data provided by the LOw Frequency ARray (LOFAR) as part of the LOFAR Two-Metre Sky Survey (LoTSS) to investigate their radio properties \cite{LoTSS}. It is a low-luminosity radio-quiet QSO. Radio continuum emission at low frequencies in low-luminosity quasars is consistent with being dominated by star formation.

2MASS J11052188+3814018 with number SDSS J125553.04+272405.2 has the spectrum obtained by the SDSS spectrograph, classified as broadline QSO. The spectrum is dominated by the QSO emission lines: broad H$\alpha$+[NII], the Balmer series -- broad H$\beta$, H$\gamma$, H$\delta$, narrow forbidden [OII], [OIII], [NeIII], [SII]. 

It was observed by XMM Newton 3 times (see Table \ref{Sample Data}, Table \ref{master}). The Observation 0124710101 was used for processing. It's spectrum is poor and it was fitted with typical model (see Pic. \ref{s1}, Table \ref{Spectr}), {\it powerlaw*phabs} with power-law index $\Gamma = 2.11 \pm 3.75 \cdot 10^{-2}$. The fit has statistical probability 59.05/45 and null hypothesis probability $7.8 \%$. The X-ray image shows the glow of the cluster's Abell 1656 halo.

2MASS J11052188+3814018 is classified as radio-quiet type I QSO. 

$\bf{SDSS J013433.43+001206.6}$ is very problematic to observe. At a distance of 113 angular seconds from the quasar there is a star HD9649 with a magnitude of V = 7.87. It blends the quasar in the optical range. The quasar is listed in the Chandra catalogs \cite{Chandra1, Chandra2}, and the supernova SN08083 was observed in it \cite{supernova}. XMM observed this area twice in 2015 as part of the Stripe 82X project \cite{Stripe}, for 5248 seconds each observation. Observation numbers are 0747420147 and 0747420148. In the previous observation, images 0747420147 has a very strong background. The quasar is lost in the image as a slight fluctuation of the background. Very few photons were registered in observation 0747420148. Among them, the quasar is detected by an automatic algorithm. The quasar's image is 30 pixels that is quite a bit. Its X-ray flux is $(1.9 \pm 0.7)\cdot 10 ^{-13} erg~s^{-1}cm^{-2}$, it is quite a lot. The probable problem is the low sensitivity of the cameras during this observation. This is one of the farthest of our quasars z = 1.391, except for the enhanced gravitational Huchra's lens. Absolute magnitude in visible spectral region (z filter) is $M=-25.7$.

In the SDSS catalog it is object number SDSS J120858.00+454035.4 with the spectrum obtained by the BOSS spectrograph, its class is Broadline QSO. The spectrum is dominated by the typical QSO emission lines: broad CIV, CIII], CII, MgII, narrow Balmer series. Besides  the  major  emission lines several weaker features are recognized as well, e.g. HeII, [OII], [NeIII]. The line content of the quasar spectrum is similar to QSO J1208+4540. 

SDSS J013433.43+001206.6 unambiguously classified as radio-quiet type I QSO.

We can divide our quasars into groups according to common features in the optical and also in X-ray spectrum. 

The first group will include radio-loud Fanaroff-Riley type I radio sources type 1 quasars: 3C189 (NGC 2484), 3C264.0 (NGC 3862), QSO B0625-354 (LEDA 2824793). The second group will include clearly similar radio-loud, type II quasar: 3C 338.0 (NGC 6166), ESO 349-10 (LEDA 73000). The third group includes radio-quiet type II QSOs: 2MASS J11052188+3814018 (LEDA 33514) (whose optical spectrum with the same lines as 3C 338.0), elusive AGN NGC 3607. The fourth group includes radio-quiet type I QSOs: QSO J1208+4540 (PG1206+459), QSO J2240+0321 (Huchra's lens), SDSS J013433.43+001206.6. Radio-quiet type I QSO 2MASS J12555303+2724053 (PB 3126) differs from the previous ones and the presence of lines in optical spectrum is more like 3C 338.0 or 2MASS J11052188+3814018 but different in wavelengths.
 
\begin{sidewaystable}[htp]
	\begin{tabularx}{\textwidth}{l@{\hspace{4pt}}*{17}{c}}\\
		\multicolumn{17}{>{\centering\setlength\hsize{1\hsize} }X}{Quasars Xgal 20}\\
		\hline
		\hline
		\multicolumn{1}{>{\centering\setlength\hsize{0.02\hsize} }X|}{N}
		&\multicolumn{1}{>{\centering\setlength\hsize{0.24\hsize} }X|}{Name $^{1}$}
		& \multicolumn{1}{>{\centering\setlength\hsize{0.1\hsize} }X|}{Alternative Name $^{2,3}$}
		& \multicolumn{1}{>{\centering\setlength\hsize{0.065\hsize} }X|}{RA (X-ray)$^3$}
		& \multicolumn{1}{>{\centering\setlength\hsize{0.065\hsize} }X|}{DEC (X-ray)$^3$}
		& \multicolumn{1}{>{\centering\setlength\hsize{0.24\hsize} }X|}{Spectral classification$^2$}
		& \multicolumn{1}{>{\centering\setlength\hsize{0.06\hsize} }X|}{Morph. type$^2$}
		& \multicolumn{1}{>{\centering\setlength\hsize{0.06\hsize} }X}{Redshift, z (Helio$^2$)}
		\\ \hline
	    \hline
		1 &3C 189 & NGC 2484 & 119,617 & 37,7865 & QSO$^1$/SyG$^1$/LERG$^2$& S0 & 0.04284  \\
		2 &3C 264.0 & NGC 3862 & 176,271 & 19,6066 & QSO$^1$/LINER$^{1,}$$^2$/LERG$^2$ & E & 0.02172\\
		3 &3C 338.0 & NGC 6166 & 247,16 & 39,5511 & QSO$^1$/LINER$^{1}$/NLRG$^2$/LEG$^2$ & cD2 pec & 0.03035 \\
		4 &ESO 349-10 & LEDA 73000 & 359,254 & -34,76 & QSO$^1$/BrClG$^2$/USS$^2$ & cD4 & 0.04905\\
		5 &NGC 3607 & LEDA 34426 & 169,228 & 18,0519 & QSO$^1$/Sy2G$^2$ & SA0 & 0.00314  \\
		6 &QSO B0625-354 & LEDA 2824793 & 96,778 & -35,4876 & QSO$^1$/BLL$^1$/Sy3G$^2$/WLRG$^2$ & E & 0.05459\\
		7 &QSO J1208+4540 & PG1206+459 & 182,242 & 45,6765 & QSO$^{1,}$$^2$ & - & 1.16494\\
		8 &QSO J2240+0321 & Huchra's Lens & 340,126 & 3,35842 & QSO$^{1,}$$^2$ & - & 1.69500 \\
		9 &2MASS J11052188+3814018 & LEDA 33514  & 166,341 & 38,2339 & QSO$^{1,}$$^2$ & - & 0.02847 \\
	    10 &2MASS J12555303+2724053  & PB 3126 & 193.971 & 38,2339 & QSO$^{1,}$$^2$/Sy1G$^1$ & - & 0.31605 \\
	    11 &SDSS J013433.43+001206.6 & --//-- & 23,6393 & 27.4014 &  QSO$^{1,}$$^2$/Extended Src$^2$ & - & 1.39268 \\
		 \hline
		\end{tabularx}
	\caption{General properties of quasars from a Xgal 20 sample.}{Name, classification and parameters for every galaxy were taken from SIMBAD$^1$, NED$^2$ databases. X-ray coordinates are obtained from the catalog of observations 4XMM-DR9$^3$. The alternative name refers here to appellation in 4XMM-DR9.}
	\label{Sample Data}

	\begin{tabularx}{\textwidth}{l@{\hspace{0.5pt}}*{17}{c}}\\
		\multicolumn{17}{>{\centering\setlength\hsize{1\hsize} }X}{The source list }\\
		\hline
		\hline
		\multicolumn{1}{>{\centering\setlength\hsize{0.02\hsize} }X|}{N}
		& \multicolumn{1}{>{\centering\setlength\hsize{0.05\hsize} }X|}{XMM
			obs.}
		& \multicolumn{1}{>{\centering\setlength\hsize{0.12\hsize} }X|}{Obs.ID}
		& \multicolumn{1}{>{\centering\setlength\hsize{0.2\hsize} }X|}{Obs.Date}
		& \multicolumn{1}{>{\centering\setlength\hsize{0.1\hsize} }X|}{Duration, s}
		& \multicolumn{1}{>{\centering\setlength\hsize{0.15\hsize} }X|}{$F, \rm erg~s^{-1}~cm^{-2}$}
		& \multicolumn{1}{>{\centering\setlength\hsize{0.15\hsize} }X|}{$\Delta F, \rm erg~s^{-1}~cm^{-2}$}
		& \multicolumn{1}{>{\centering\setlength\hsize{0.07\hsize} }X}{Spectra}
				\\ \hline
		\hline
		1 & 1 & 0602390101 & 27.04.2010 2:21:56 & 69052& 3.9532$\cdot 10^{-13}$ & 9.75638$\cdot 10^{-15}$ & True\\
		2 & 3 & 0061740101 & 26.05.2001 19:49:25 & 33208& 2.46508$\cdot 10^{-12}$ & 6.35938$\cdot 10^{-14}$ & True\\
		3 & 5 & 0723801101 & 12.08.2013 20:52:39 & 57000& 5.82564$\cdot 10^{-11}$ & 3.39312$\cdot 10^{-13}$ & True\\
		4 & 4 & 0723801001 & 13.05.2013 13:52:04 & 97783& 2.22306$\cdot 10^{-11}$ & 1.62023$\cdot 10^{-13}$  & True\\
		5 & 2 & 0693300101 & 20.05.2012 1:54:39 & 58770& 3.8614$\cdot 10^{-13}$ & 4.00185$\cdot 10^{-14}$ & True\\
		6 & 3 & 0302440601 & 27.08.2005 23:18:32 & 13809& 8.18398$\cdot 10^{-12}$ & 7.05082$\cdot 10^{-14}$ & True \\
		7 & 1 & 0033540601 & 11.05.2002 4:50:46 & 13190& 4.04372$\cdot 10^{-13}$ & 3.25894$\cdot 10^{-14}$ & True \\
		8 & 3 & 0823730101 & 19.05.2018 21:18:47 & 141600& 2.66877$\cdot 10^{-13}$ & 5.57999$\cdot 10^{-15}$ & True \\
		9 & 131 & 0153950701 & 05.05.2002 3:51:30 & 19982& 1.69502$\cdot 10^{-13}$ & 2.60988$\cdot 10^{-14}$ & True\\
		10 & 3 & 0124710101 & 21.06.2000 8:25:20 & 41505& 1.05904$\cdot 10^{-12}$ & 4.56635$\cdot 10^{-14}$ & True \\
		11 & 2 & 0747420147 & 28.01.2015 11:26:56 & 5247& 1.1499$\cdot 10^{-13}$ & 1.11965$\cdot 10^{-13}$ & False \\
		\hline
	\end{tabularx}
	\caption{Xmm-Newton observations of quasars from a Xgal 20 sample.}{The parameters are obtained from the catalog of observations 4XMM-DR9 for the EPIC PN camera. Rows retrieved from xmmmaster.}
	\label{master}
\end{sidewaystable}

\begin{sidewaystable}[htp]
\begin{tabularx}{\textwidth}{l@{\hspace{0.5pt}}*{17}{c}}\\
	\multicolumn{17}{>{\centering\setlength\hsize{1\hsize} }X}{The source list}\\
	\hline
	\hline
	\multicolumn{1}{>{\centering\setlength\hsize{0.02\hsize} }X|}{N}
	& \multicolumn{1}{>{\centering\setlength\hsize{0.2\hsize} }X|}{Name}
	& \multicolumn{1}{>{\centering\setlength\hsize{0.11\hsize} }X|}{HR1}
	& \multicolumn{1}{>{\centering\setlength\hsize{0.11\hsize} }X|}{$\Delta$HR1}
	& \multicolumn{1}{>{\centering\setlength\hsize{0.11\hsize} }X|}{HR2}
	& \multicolumn{1}{>{\centering\setlength\hsize{0.11\hsize} }X|}{$\Delta$HR2}
	& \multicolumn{1}{>{\centering\setlength\hsize{0.11\hsize} }X|}{HR3}
	& \multicolumn{1}{>{\centering\setlength\hsize{0.11\hsize} }X}{$\Delta$HR3}
	\\ \hline
	\hline
	1 & NGC 2484 & 0.37363& 0.0153193 & -0.26219& 0.0153231& -0.508318 &0.020391\\
	2 & NGC 3862 & 0.019719& 0.0127665 & -0.260278& 0.0139571& -0.498934 &0.018638\\
	3 & NGC 6166 & 0.23734& 0.0048333 & 0.0310428& 0.00227425& -0.46616 &0.00458736\\
	4 & LEDA 73000 & 0.330147& 0.00754817 & 0.0594553& 0.0066729& -0.509421 &0.00644842\\
	5 & LEDA 34426 & 0.399366& 0.0324467 & -0.415228& 0.0324575& -0.427094 &0.0566647\\
	6 & LEDA 2824793 & 0.223355 & 0.00645387 &-0.160546& 0.00641356& -0.523812 &0.00760857\\
	7 & PG1206+459 & 0.0685034 & 0.0526117 & 0.0407695& 0.0501199& -0.399961 &0.0581317\\
	8 & Huchra's Lens & 0.263522& 0.0181575 & 0.122163& 0.0152941& -0.240262 &0.0159737\\
	9 & LEDA 33514 & 0.538954& 0.0975307 & -0.426381 & 0.103274& -0.224261 &0.170275\\
	10 & PB 3126 & -0.0209306& 0.0226295 & -0.110186 & 0.024142& -0.442848 &0.0283414\\
	11 & SDSS J013433.43+001206.6 & 0.605304& 0.200941 & -0.471439& 0.255287& -0.0709096 &0.521749\\
	\hline
\end{tabularx}
\caption{Xmm-Newton observations and three hardness ratios.}{The parameters are obtained from the catalog of observations 4XMM-DR9 for the EPIC PN camera.}
\label{hr}

\begin{tabularx}{\textwidth}{l@{\hspace{4pt}}*{17}{c}}\\
	\multicolumn{17}{>{\centering\setlength\hsize{1\hsize} }X}{Spectral fitting results for the 10 brighter sources}\\
	\hline
	\hline
	\multicolumn{1}{>{\centering\setlength\hsize{0.02\hsize} }X|}{N}
	& \multicolumn{1}{>{\centering\setlength\hsize{0.15\hsize} }X|}{Photon Index, $\Gamma$}
	& \multicolumn{1}{>{\centering\setlength\hsize{0.16\hsize} }X|}{$nH,~10^{22}~cm^{-2}$}
	& \multicolumn{1}{>{\centering\setlength\hsize{0.16\hsize} }X|}{kT, (apec/mekal)~keV}
	& \multicolumn{1}{>{\centering\setlength\hsize{0.15\hsize} }X|}{kT, (bbody)~keV}
	& \multicolumn{1}{>{\centering\setlength\hsize{0.15\hsize} }X|}{Line E, (zgauss)~keV}
	& \multicolumn{1}{>{\centering\setlength\hsize{0.06\hsize} }X}{$\chi^2$/d.o.f.}
	
	\\ \hline
	\hline
	1 & $2.396\pm4.5\cdot 10^{-2}$& 0.100 frozen &$0.730\pm3.9\cdot 10^{-2}$ & -- &  -- & 64.43/58  \\
	2 &$2.542 \pm 7.7\cdot 10^{-2}$ & $3.3\cdot 10^{-2}\pm1.1\cdot 10^{-2}$ & -- & -- & -- & 86.02/83\\
	3 &$2.209\pm2.8\cdot 10^{-2}$ & $0.120\pm5\cdot 10^{-3}$ & $3.016\pm5.0\cdot 10^{-2}$ & -- & 6.500 frozen & 961.27/740 \\
	4 &$2.214 \pm 5.0\cdot 10^{-2}$ & $0.123 \pm 1.0\cdot 10^{-2}$ & $1.844 \pm 7.0\cdot 10^{-2}$ & -- & $6.726 \pm 3.1\cdot 10^{-2}$ & 351.16/296 \\
	5 &-- & -- & -- & $0.237\pm1.2\cdot 10^{-2}$ & -- & 10.6/6  \\
	6 &$2.496 \pm 2.3\cdot 10^{-2}$ & $8.3\cdot 10^{-2} \pm 3\cdot 10^{-3}$ & -- & -- & -- & 367.82/308\\
	7 &$1.842 \pm 0.233$ & $4.58\cdot 10^{-2} \pm 4.55\cdot 10^{-2}$ & -- & -- & -- & 11.70/10\\
	8 &$1.509 \pm 5.8\cdot 10^{-2}$ & $0.497 \pm 6.1\cdot 10^{-2}$ & 1.000 frozen & 1.000 frozen & -- & 119.05/101\\
	9 &-- & --  &$ 0.649 \pm 2.5\cdot 10^{-2}$ &$ 0.129 \pm 1.9\cdot 10^{-2}$ & -- & 40.50/30 \\
	10 &$2.114\pm 3.8\cdot 10^{-2}$ & 1.0$\cdot 10^{-2}$ frozen & -- & -- & -- & 59.05/45 \\
	\hline
\end{tabularx}
\caption{Xmm Newton observing and fitting their spectrum with the Xspec models.}{The data extraction was performed with the Science Analysis System (SAS).}
\label{Spectr}

\end{sidewaystable}

\begin{figure}[!htb]
	\minipage{0.49\textwidth}
	\includegraphics[width=\linewidth]{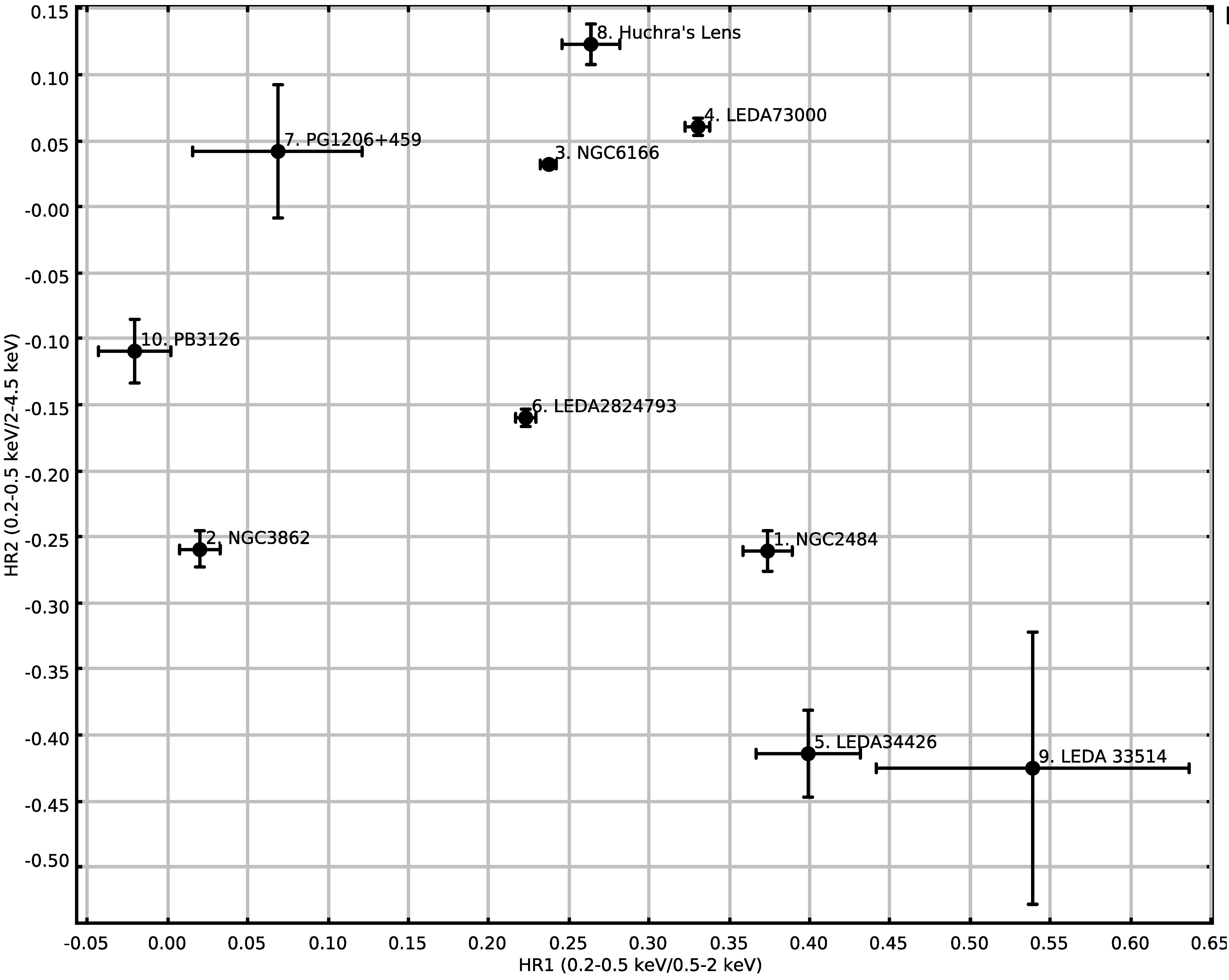}
	\caption{The HR2 (0.5-4.5 keV) as a function of 
		the HR1 (0.2-2 keV) hardness ratios for quasars Xgal 20.}\label{hr12}
	\endminipage
	\hfill
	\minipage{0.49\textwidth}
	\includegraphics[width=\linewidth]{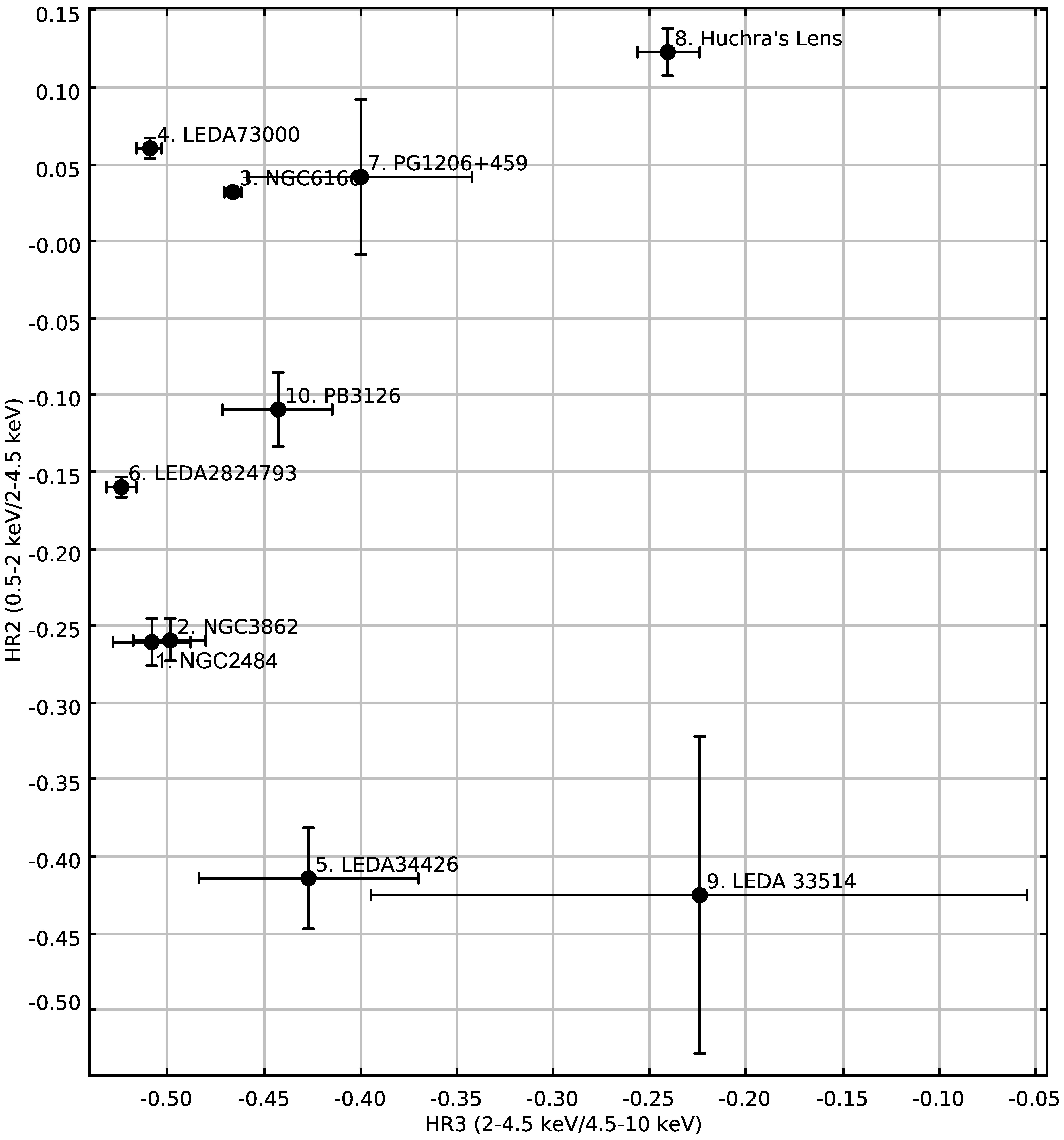}
	\caption{The HR2 (0.5-4.5 keV) as a function of 
		the HR3 (2-8 keV) hardness ratios for quasars Xgal 20.}\label{hr23}
	\endminipage
\end{figure}

It is interesting that at a short distance from the observer, we got quite a lot of type II QSOs (36 percent among all QSOs in our sample), while at large distances, it is naturally, more difficult to search for type II quasars, as if its obscured. Regardless of the dependence of an increase in the number of quasars on the redshift, it is obvious, that there are great prerequisites to search and study obscured quasars closer to the observer.

With some caution, we would call our classification preliminary for only a few QSOs, precisely because of insufficient data on some objects with less reliable type II identification. And they should have been observed in more detail. 

\subsection{The Hardness Ratios}

\indent \indent We obtained three hardness ratios using the catalog of observations 4XMM-DR9 in the following energy bands 
S1=0.2-0.5 keV, S2=0.5-2 keV, M=2-4.5 keV and H=4.5-8 keV.
The hardness ratios are defined as 
HR1=(S2-S1)/(S2+S1), HR2=(M-S2)/(M+S2), HR3=(H-M)/(H+M).

The hardness ratios are given in Table \ref{hr} together with their $1\sigma$ uncertainties.  
As seen from the energy differences of each HR, HR1 is more sensitive    
to the presence of a soft excess component or  
a low amounts of column density ($> 10^{20}~\rm cm^{-2} $).
HR2 is affected by the presence of $\sim10^{21}~\rm cm^{-2} $, while HR3 values could be sensitive to a reflection component or to even larger columns ($> 10^{22}~\rm cm^{-2}$) \cite{Akylas}.   
In Fig. \ref{hr12} we plot the HR2 versus HR1 for our 10 sources and in Fig. \ref{hr23} we plot the HR2 versus the HR3 hardness ratio. 

In Fig. \ref{hr12} we find one source 2MASS J11052188+3814018 (LEDA 33514) with a spectrum harder than $\rm HR1\sim 0.5$ in the soft band. It should have softer spectra in the hard (2-8 keV) band (see the HR3 values Table \ref{hr}), but the spectrum that we fit is cut off by 2 keV. The closest source NGC 3607 (LEDA 34426) to the above in the diagram is also a type II radio-quiet QSO with a spectrum that is cut off by soft X-rays. 
The radio-quiet quasar QSO J1208+4540 (PG 1206+459) is the closest to Huchra's lens at Fig. \ref{hr12}. Both quasars have hard spectra (photon indexes 1.84 and 1.51 correspondingly), which agrees with arbitrary X-ray QSO parameters studied by \cite{Fiore, Akiyama}.

In general, observations indicate that the X-ray power law index spans a limited range, roughly between 1.5 and 2.5 \cite{Ishibashi, Shemmer6, Shemmer8}. The power law spectrum is harder in massive and luminous quasars compared to less massive sources. Thus the trend indicating a steepening of the observed spectrum for lower black hole mass and higher accretion rates is confirmed for the Palomar-Green (PG) quasar sample \cite{Piconcelli}. Very steep spectra ($\sim 2.3$) are expected for low-mass objects accreting at high accretion rates. Extremely steep spectra are in fact observed in the particular class of Narrow-line Seyfert 1 (NLS 1) galaxies, which are believed to be powered by small black holes accreting at rates close to the Eddington limit \cite{Ishibashi}. 
Extremely steep photon index ($\sim 2.4 - 2.5$) in our sample occurs in three sources -- radio-loud type I quasars. The first two 3C 189 (NGC 2484), 3C 264.0 (NGC 3862) with broad absorption and narrow forbidden lines in the spectrum, and the third source is QSO B0625-354 (LEDA 2824793) low excitation line AGN transitional Fanaroff-Riley type I/BL Lac object, source of gamma radiation.     

\section {Conclusions}
\indent \indent We present results of a systematic analysis of the XMM-Newton spectra of nearby optically bright QSOs. The objects have been selected from the Xgal20 sample.

We characterize the X-ray spectral properties of optically selected QSOs in the 0.1 -- 10 keV energy band (soft and hard X-rays). A power law component with absorbtion accounts the hard band spectra with an average photon index of $2.17\pm 0.07$. In some cases it is able to account the soft excess by single temperature black-body model. Two quasars fall significantly out of statistics because X-ray spectra of these objects show strong soft excess below 2 keV and at higher energies, the spectrum was cut off and/or was absorbed. The presence of a complex absorption/emission pattern in the soft X-ray band manifests in 5 objects. In two QSOs we detected FeK$\alpha$ line. 

We consider as an advantage, that the models were selected by statistical method so that the result had the best probability. 

The main result of our work is that for each of considered quasars the type (type I or type II) was established both based on the observed X-ray spectrum and analyzing previous other authors works, data and optical spectra. 

\section {Acknowledgment}

\indent \indent This research has made with support of the Center for the Collective Use of Scientific Equipment "Laboratory of high energy physics and astrophysics". 

This research has made use of data obtained from the 4XMM XMM-Newton serendipitous source catalogue compiled by the 10 institutes of the XMM-Newton Survey Science Centre selected by ESA.

We acknowledge the usage of the HyperLeda database. 

This research has made use of the NASA/IPAC Extragalactic Database (NED),
which is operated by the Jet Propulsion Laboratory, California Institute of Technology,
under contract with the National Aeronautics and Space Administration.

This research has made use of the SIMBAD database, operated at CDS, Strasbourg, France. 

Funding for the Sloan Digital Sky 
Survey IV has been provided by the 
Alfred P. Sloan Foundation, the U.S. 
Department of Energy Office of 
Science, and the Participating 
Institutions. SDSS-IV acknowledges support and 
resources from the Center for High 
Perfor-mance Computing  at the 
University of Utah. The SDSS 
website is www.sdss.org.

\renewcommand{\refname}

\end{document}